\pdfoutput=1

\documentclass[camera,letterpaper,nomarginnotes,nonarrowgutter]{jpaper}

\usepackage{algorithmic}
\usepackage{array}
\usepackage{stfloats}

\usepackage{cite}
\usepackage{amsmath,amssymb,amsfonts}
\usepackage{graphicx}
\usepackage{textcomp}
\usepackage{xcolor}
\usepackage{microtype}
\usepackage{url}
\usepackage[T1]{fontenc}
\usepackage{booktabs}
\usepackage{pgfplots}
\usepackage{pgfplotstable}
\usepackage{todonotes}
\usepackage{xargs}
\usepackage{balance}
\usepackage{multirow,booktabs,siunitx,graphicx}
\usepackage{fancyhdr}
\usepackage{duckuments}
\usepackage{tabularx}
\usepackage{subcaption}

\pgfplotsset{compat=1.18}
\usepackage{tikz}
\usetikzlibrary{patterns, arrows.meta, positioning, shapes.geometric}

\usepackage[acronym,nonumberlist,nowarn]{glossaries}
\glsdisablehyper
\newacronym{ADI}{ADI}{Alternating Direction Implicit}
\newacronym{AGU}{AGU}{Address Generation Unit}
\newacronym[longplural={Access History Tables}]{AHT}{AHT}{Access History Table}
\newacronym{AHT-R}{AHT-R}{Access History Table - Read}
\newacronym{AHT-W}{AHT-W}{Access History Table - Write-Back}
\newacronym{AIP}{AIP}{Access Interval Predictor}
\newacronym{TPOT}{TPOT}{time per output token}
\newacronym{SLO}{SLO}{service-level objective}

\newacronym{AL}{AL}{Added Latency for column accesses}
\newacronym{ASIC}{ASIC}{Application Specific Integrated Circuit}
\newacronym{AVX}{AVX}{Advanced Vector Extensions}
\newacronym{FP8}{FP8}{8-bit floating-point}
\newacronym{FP4}{FP4}{4-bit floating-point}
\newacronym{BF16}{BF16}{16-bit brain floating-point}
\newacronym[longplural={Basic Block Vectors}]{BBV}{BBV}{Basic Block Vector}
\newacronym{BHT}{BHT}{Branch History Table}
\newacronym{ECC}{ECC}{error correction code}
\newacronym{MABM}{MABM}{mat-aware bit mapping}
\newacronym{HBM}{HBM}{high bandwidth memory}
\newacronym{SLC}{SLC}{single-level cell}
\newacronym{GPU}{GPU}{graphics processing unit}
\newacronym{TPU}{TPU}{tensor processing unit}
\newacronym{NPU}{NPU}{neural processing unit}
\newacronym{HBF}{HBF}{high bandwidth flash}
\newacronym{D2D}{D2D}{die-to-die}
\newacronym{CBA}{CBA}{CMOS directly bonded to array}

\newacronym{DDR4}{DDR4}{Double-Data Rate 4}
\newacronym{MIMD}{MIMD}{multiple-instruction multiple-data}
\newacronym{SIMT}{SIMT}{Single Instruction Multiple Threads}
\newacronym{SLP}{SALP}{subarray-level parallelism}
\newacronym{BLP}{BLP}{bank-level parallelism}
\newacronym{LLM}{LLM}{large language model}

\newacronym{NN}{NN}{neural network}
\newacronym{BTB}{BTB}{Branch Target Buffer}
\newacronym{CAS}{CAS}{Column Address Strobe}
\newacronym{AMAT}{AMAT}{Average Memory Access Time}
\newacronym{CCD}{CCD}{Column to Column Delay}
\newacronym{AI}{AI}{Arithmetic Intensity}
\newacronym[longplural={Columnar Database Systems}]{CDBS}{CDBS}{Columnar Database System}
\newacronym[longplural={Chip Multiprocessors}]{CMP}{CMP}{Chip Multiprocessor}
\newacronym{CMT}{CMT}{Chip Multithreading}
\newacronym[longplural={Coarse-Grain Reconfigurable Arrays}]{CGRA}{CGRA}{Coarse-Grain Reconfigurable Array}
\newacronym{C-RAM}{C-RAM}{Computational-RAM}
\newacronym{CWD}{CWD}{Column Write Delay}
\newacronym{DBI}{DBI}{Dynamic Binary Instrumentation}
\newacronym[longplural={Database Management Systems}]{DBMS}{DBMS}{Database Management System}
\newacronym{DDR}{DDR}{Double Data Rate}
\newacronym{DEWP}{DEWP}{Dead Line and Early Write-Back Predictor}
\newacronym{DRAM}{DRAM}{Dynamic Random Access Memory}
\newacronym{DSBP}{DSBP}{Dead Sub-Block Predictor}
\newacronym{DSM}{DSM}{Decomposed Storage Manage}
\newacronym{FAW}{FAW}{Four row Activation Window}
\newacronym{FP}{FP}{Floating-Point}
\newacronym{EDP}{EDP}{Energy-Delay Product}
\newacronym{FCFS}{FCFS}{First-Come First-Serve}
\newacronym{FIFO}{FIFO}{first-in first-out}
\newacronym{FSM}{FSM}{finite state machine}
\newacronym{FPGA}{FPGA}{Field-Programmable Gate Array}
\newacronym[longplural={Functional Units}]{FU}{FU}{Functional Unit}
\newacronym{GAg}{GAg}{Global Adaptive branch prediction using one Global PHT}
\newacronym{GAs}{GAs}{Global Adaptive branch prediction using per-Set PHT}
\newacronym{GCC}{GCC}{GNU Compiler Collection}
\newacronym{GEMS}{GEMS}{General Execution-driven Multiprocessor Simulator}
\newacronym[longplural={General Purpose Processors}]{GPP}{GPP}{General Purpose Processor}
\newacronym{HMC}{HMC}{Hybrid Memory Cube}
\newacronym{HIVE}{HIVE}{HMC Instruction Vector Extensions}
\newacronym{IATAC}{IATAC}{Inter-Access Time per Access Count}
\newacronym{ILP}{ILP}{Instruction Level Parallelism}
\newacronym{IPC}{IPC}{Instructions per Cycle}
\newacronym{ISA}{ISA}{Instruction Set Architecture}
\newacronym{IRAM}{IRAM}{Intelligent RAM}
\newacronym{KIPS}{KIPS}{Kilo Instructions per Second}
\newacronym{LDS}{LDS}{Linked Data Structure}
\newacronym{LFSR}{LFSR}{Linear Feedback Shift Register}
\newacronym{LFMR}{LFMR}{Last-to-First Miss-Ratio}
\newacronym{MLP}{MLP}{Memory-Level Parallelism}

\newacronym{LLC}{LLC}{last-level cache}
\newacronym{LRU}{LRU}{Least Recently Used}
\newacronym{LSU}{LSU}{Load Store Unit}
\newacronym{LTP}{LTP}{Last-Touch Predictor}
\newacronym{LvP}{LvP}{Live-time Predictor}
\newacronym{LWP}{LWP}{Last Write Predictor}
\newacronym{MARSS}{MARSS}{Micro Architectural and System Simulator}
\newacronym{McPAT}{McPAT}{Multi-core Power, Area, and Timing}
\newacronym{MIPS}{MIPS}{Microprocessor without Interlocked Pipeline Stages}
\newacronym{MOB}{MOB}{Memory Order Buffer}
\newacronym{MMU}{MMU}{memory management unit}
\newacronym{PuD}{PUD}{processing-using-DRAM}
\newacronym{MPKI}{MPKI}{Misses per Kilo-Instruction}
\newacronym{MSHR}{MSHR}{Miss-Status Handling Registers}
\newacronym{NAS}{NAS}{Numerical Aerodynamic Simulation}
\newacronym{NDP}{NDP}{Near-Data Processing}
\newacronym{NMP}{NMP}{Near Memory Processor}
\newacronym{NoC}{NoC}{Network-on-Chip}
\newacronym{NPB}{NPB}{NAS Parallel Benchmark}
\newacronym{NUCA}{NUCA}{Non-Uniform Cache Architecture}
\newacronym{NUMA}{NUMA}{Non-Uniform Memory Access}
\newacronym{OoO}{OoO}{Out-of-Order}
\newacronym{OpenMP}{OpenMP}{Open Multi-Processing}
\newacronym{OS}{OS}{Operating System}
\newacronym{PAg}{PAg}{Per-address Adaptive branch prediction using one Global PHT}
\newacronym{PAs}{PAs}{Per-address Adaptive branch prediction using per-Set PHT}
\newacronym{PC}{PC}{Program Counter}
\newacronym[longplural={Pointer-Chasing Engines}]{PCE}{PCE}{Pointer-Chasing Engine}
\newacronym{PCM}{PCM}{Performance Counter Monitor}
\newacronym{PIM}{PIM}{processing-in-memory}
\newacronym{FTL}{FTL}{flash translation layer}
\newacronym{SSD}{SSD}{solid-state drive}
\newacronym{MoE}{MoE}{mixture-of-experts}

\newacronym[longplural={Pattern History Tables}]{PHT}{PHT}{Pattern History Table}
\newacronym{RAPL}{RAPL}{Running Average Power Limit}
\newacronym{RAS}{RAS}{Row Address Strobe}
\newacronym{RAT}{RAT}{Registers Alias Table}
\newacronym{RC}{RC}{Row Cycle}
\newacronym{RCD}{RCD}{RAS to CAS Delay}
\newacronym[longplural={Row-based Database Systems}]{RDBS}{RDBS}{Row-based Database System}
\newacronym{ROB}{ROB}{Reorder Buffer}
\newacronym{RRD}{RRD}{Row to Row activation Delay}
\newacronym{DNN}{DNN}{deep neural network}
\newacronym{ANN}{ANN}{Artificial Neural Network}
\newacronym{PnM}{PNM}{processing-near-memory}
\newacronym{PuM}{PUM}{processing-using-memory}
\newacronym{FFD}{FFD}{first-fit-decreasing}
\newacronym{HFF}{HFF}{helper flip-flop}
\newacronym{OBPS}{OBPS}{one-bit per-subarray}
\newacronym{ABOS}{ABOS}{all-bits in one-subarray}
\newacronym{ABPS}{APBS}{all-bits per-subarray}
\newacronym{GEMV}{GEMV}{general matrix-vector product}

\newacronym{RBR}{RBR}{redundant binary representation}
\newacronym{RP}{RP}{Row Precharge}
\newacronym{RTL}{RTL}{Register Transfer Level}
\newacronym{RTP}{RTP}{Read To Precharge}
\newacronym{RVU}{RVU}{Reconfigurable Vector Unit}
\newacronym{SDP}{SDP}{Skewed Dead-Block Predictor}
\newacronym{SESC}{SESC}{Superescalar Simulator}
\newacronym{SSE}{SSE}{Streaming SIMD Extensions}
\newacronym{SFP}{SFP}{Spatial Footprint Predictor}
\newacronym{SiNUCA}{SiNUCA}{Simulator of Non-Uniform Cache Architectures}
\newacronym{SMT}{SMT}{Simultaneous Multi-Threading}
\newacronym{SIMD}{SIMD}{single-instruction multiple-data}
\newacronym{LUT}{LUT}{lookup table}
\newacronym{SM}{SM}{streaming multiprocessors}
\newacronym{LHB}{LHB}{latency-hiding buffer}

\newacronym{SPEC}{SPEC}{Standard Performance Evaluation Corporation}
\newacronym{SoC}{SoC}{System-on-Chip}
\newacronym{SPP}{SPP}{Spatial Pattern Predictor}
\newacronym{SRAM}{SRAM}{Static Random Access Memory}
\newacronym{SSOR}{SSOR}{Symmetric Successive Over-Relaxation}
\newacronym{SSV}{SSV}{Search Set Vector}
\newacronym{TLB}{TLB}{Translation Look-aside Buffer}
\newacronym{TLP}{TLP}{Thread Level Parallelism}
\newacronym[longplural={through-silicon vias}]{TSV}{TSV}{through-silicon via}
\newacronym{VWQ}{VWQ}{Virtual Write Queue}
\newacronym{WTR}{WTR}{Write Recovery time}
\newacronym{WR}{WR}{Write To Read delay time}
\newacronym{TRA}{TRA}{triple row activation}

\usepackage{xspace}
\usepackage[binary-units=true,per-mode=symbol]{siunitx}

\usepackage{cleveref}
\crefformat{section}{\S#2#1#3}
\crefformat{subsection}{\S#2#1#3}
\crefformat{subsubsection}{\S#2#1#3}

\usepackage{enumitem}
\usepackage{marvosym}
\usepackage{pifont}

\newif\ifsubmission
\submissiontrue

\ifsubmission
    \newcommand{\gf}[1]{#1}
    \newcommand{\gfhpca}[1]{#1}
    \newcommand{\acy}[1]{#1}
    \newcommand{\gfb}[1]{}

    \newcommandx{\feedback}[2][1=]{}
    \newcommandx{\change}[2][1=]{}

    \newcommandx{\info}[2][1=]{}

\else    
    \paperwidth=\dimexpr\paperwidth + 4cm\relax
    \oddsidemargin=\dimexpr\oddsidemargin + 2cm\relax
    \evensidemargin=\dimexpr\evensidemargin + 2cm\relax
    \marginparwidth=\dimexpr\marginparwidth + 2cm\relax

    \newcommandx{\gfb}[1]{\textcolor{blue}{\textit{GF: #1}}}
    \newcommand{\gf}[1]{\textcolor{blue}{#1}}
    \newcommand{\gfhpca}[1]{\textcolor{red}{#1}}
    \newcommand{\acy}[1]{\textcolor{green}{#1}}
    
     \newcommandx{\feedback}[2][1=]{\todo[linecolor=blue,backgroundcolor=blue!25,bordercolor=blue,#1,size=\tiny]{#2}}
     \newcommandx{\change}[2][1=]{\todo[linecolor=red,backgroundcolor=red!25,bordercolor=red,#1,size=\tiny]{#2}}
    \newcommandx{\info}[2][1=]{\todo[linecolor=yellow,backgroundcolor=yellow!25,bordercolor=yellow,#1,size=\tiny]{#2}}
\fi

\newcommand{\paratitle}[1]{\vspace{4pt}\noindent\textbf{#1.}}

\newcommand{\li}{(\textit{i})}
\newcommand{\lii}{(\textit{ii})}
\newcommand{\liii}{(\textit{iii})}
\newcommand{\liv}{(\textit{iv})}
\newcommand{\lv}{(\textit{v})}

\DeclareSIUnit{\billion}{B}
\DeclareSIUnit{\trillion}{T}

\newcommand{\prop}{FLINT\xspace}
\newcommand{\propLong}{\underline{fl}ash \underline{in}ference \underline{t}ier\xspace}

\newcommand{\circled}[1]{\tikz[baseline=(char.base)]{\node[shape=circle,draw,inner sep=0pt,fill=black, text=white] (char) {#1};}}

\newcommand{\revdel}[1]{}
\newcommand{\revdelhpca}[1]{}

\makeatletter
\def\bstctlcite{\@ifnextchar[{\@bstctlcite}{\@bstctlcite[@auxout]}}
\def\@bstctlcite[#1]#2{\@bsphack
  \@for\@citeb:=#2\do{%
    \edef\@citeb{\expandafter\@firstofone\@citeb}%
    \if@filesw\immediate\write\csname #1\endcsname{\string\citation{\@citeb}}\fi}%
  \@esphack}
\makeatother 

\begin{document}
\bstctlcite{IEEEexample:BSTcontrol}

\title{\scalebox{1}{\prop: Efficiently Leveraging High Bandwidth Flash}\\ \scalebox{1}{for Capacity-Scalable LLM Inference Acceleration}}

\author{
*Geraldo F. Oliveira$^\dagger$~\qquad  
*Arash Tavakkol$^\dagger$~\qquad  
Xiangyu Zhu$^\dagger$~\qquad 
Ahmet Caner Yüzügüler$^\dagger$~\qquad \\ 
Vamanan Arulchelvan$^\dagger$~\qquad
Lukas Cavigelli$^\dagger$~\qquad 
Renzo Andri$^\dagger$~\qquad 
Mohammad Sadrosadati$^\dagger$~\qquad \\ 
Jia Xinglei$^\ddagger$~\qquad   
Onur Mutlu\textsuperscript{\S}~\qquad 
Zhou Ke$^\nabla$~\qquad  
Shai Bergman$^\dagger$~\qquad  
Ji Zhang$^\ddagger$\\ \\
$^\dagger$~\emph{Huawei Technologies Switzerland AG} \quad  
$^\ddagger$~\emph{Huawei Technologies Co., Ltd.} \quad
\textsuperscript{\S}~\emph{ETH Zürich} \quad
$^\nabla$~\emph{HUST}
}

\pagenumbering{arabic}
\renewcommand{\headrulewidth}{0pt}
\fancyhf{} 
\fancyfoot[C]{\textbf{\thepage}} 
\pagestyle{plain}

\maketitle

\begingroup
\renewcommand{\thefootnote}{}
\footnotetext{*Geraldo F. Oliveira and Arash Tavakkol are co-primary authors.}
\endgroup

\setcounter{footnote}{0}
\renewcommand{\thefootnote}{\arabic{footnote}}

\begin{abstract}

\gf{\Gls{LLM} inference is increasingly constrained by accelerator
memory capacity rather than compute throughput.
This constraint is especially acute in single-accelerator and small-node inference systems, where limited on-package memory capacity restricts the size of deployable models.
\Gls{HBF} is an emerging 3D-stacked NAND flash technology that provides
multi-terabyte near-accelerator capacity, making it a promising capacity tier
for storing \gls{LLM} weights.
However, existing \gls{HBF}-based proposals face three adoption challenges: 
they 
\li~rely on coarse-grained static prefetching \gfhpca{for \gls{LLM} weights aiming to hide the $\mu$$s$-level read latency of the NAND flash device while maximizing \gls{HBF}'s read throughput},
\lii~expose NAND flash \gfhpca{management tasks (e.g., refresh operations)} to the accelerator-visible critical \gfhpca{inference} path, and
\liii~\gfhpca{miss optimization opportunities to specialize and optimize the} flash-management mechanisms to \gfhpca{the workload behavior}.
}

\gf{Our \textbf{goal} is to design an efficient \gls{HBF} substrate that integrates \gls{HBF} as a memory-capacity tier alongside \gls{HBM} while addressing these three challenges.
To this end, we propose \prop (\propLong), a workload-driven \gls{HBF} substrate for capacity-scalable \gls{LLM} inference.
\revdelhpca{\prop exploits the key observation that, during \gls{LLM} inference, model weights are \acy{read-only}, consumed in predictable layer-by-layer order, and accessed as long contiguous memory regions that naturally form \emph{burst-sized} memory transfers.
Based on this observation,} \prop introduces three mechanisms:
\li~a hardware \emph{burst-buffer controller} that \gfhpca{\emph{dynamically} coalesces and} pipelines \gls{HBF} reads \gfhpca{aiming to utilize existing NAND flash buffers while sustaining high \gls{HBF} bandwidth},
\lii~a \emph{phantom-plane refresh} mechanism, which removes refresh from the critical \gfhpca{inference} path \gfhpca{by moving refresh-related NAND flash operations outside the read foreground back via low-cost resource duplication}, and
\liii~a \emph{read-only \gls{FTL}}, which replaces \gls{SSD}-class support for arbitrary writes with a compact table that translates logical weight bursts to physical \gls{HBF} locations.}

\gf{We evaluate \prop across dense and \gls{MoE} \gls{LLM} inference
models.
\gfhpca{Our evaluation shows that \prop improves decode throughput by
1{,}205$\times$, 2.2$\times$, and 6.2$\times$, and reduces energy
consumption by 408$\times$, 1.1$\times$, and 6.8$\times$ that of an
\gls{SSD}-equipped GPU system, an \gls{HBM}-only {GPU} system, and a
prior hybrid \gls{HBM}+\gls{HBF} {GPU} system, respectively.
\prop meets a \SI{50}{\milli\second}  \gls{TPOT} \gls{SLO} with $3.1\times$ fewer GPU
packages than the \gls{HBM}-only GPU system.
\prop adds small area cost to an \gls{HBF} die (\SI{3.1}{\percent})
and \gls{HBF} base die (\SI{3.9}{\milli\meter\squared} at
\SI{7}{\nano\meter}).}
}
\end{abstract}

\glsresetall

\section{Introduction}
\label{sec:intro}

\gf{The progress of \glspl{LLM}~\cite{vaswani2017attention,brown2020language,achiam2023gpt4,dubey2024llama3,kaplan2020scaling,deepseekai2026deepseekv4,liu2024deepseek} is increasingly shaped by a widening mismatch between model size and accelerator memory capacity~\cite{gholami2024airadar,sevilla2022compute,pope2023efficiently,kwon2023efficient}.
On the algorithmic side, parameter counts have grown from hundreds of
millions in early transformer-based models~\cite{vaswani2017attention,
devlin2018bert} to hundreds of billions \gfhpca{and, most recently, trillions in frontier models such as Qwen3-235B-A22B~\cite{bai2025qwen3} (\SI{235}{\billion} total parameters), 
Llama-4 Maverick~\cite{meta2025llama} (\SI{400}{\billion}), 
Llama~3.1-405B~\cite{dubey2024llama3} (\SI{405}{\billion}), 
DeepSeek-V3~\cite{liu2024deepseek} (\SI{671}{\billion}),
Kimi~K2~\cite{team2025kimi} (\SI{1}{\trillion}), and
DeepSeek V4~\cite{deepseekai2026deepseekv4} (\SI{1.6}{\trillion}).}
\revdelhpca{This growth has yielded substantial improvements in reasoning~\cite{wei2022chain,guo2025deepseek,jaech2024openai}, instruction-following~\cite{ouyang2022training,wei2021finetuned}, and multimodal capabilities~\cite{liu2023visual, team2023gemini}, with little evidence that the scaling pressure is abating~\cite{hoffmann2022chinchilla, meta2025llama4}.}
At inference time, this parameter growth becomes a \emph{memory-capacity problem}, since serving a model requires keeping its weights resident near the accelerator, and weight capacity scales with parameter count.
\gfhpca{For example, in \gls{BF16} precision~\cite{jouppi2020domain}, model weights require two bytes per parameter, so weights alone require \gfhpca{on-package storage capacity of} roughly \SI{1.3}{\tera\byte} for
DeepSeek-V3 and \SI{810}{\giga\byte} for Llama~3.1-405B.
Recent frontier models cut the storage cost of each parameter by shipping in
narrow bit-width data formats, such as the \gls{FP8}~\cite{micikevicius2022fp8} weights of Llama-4 Maverick and Kimi~K2, or mixed data formats, such as the \gls{FP8} weights with \gls{FP4}~\cite{rouhani2023microscaling} experts in DeepSeek-V4-Pro.
However, narrow bit-width data formats do \emph{not} remove the capacity requirement: Kimi~K2 still requires \SI{1.0}{\tera\byte} of weights at its native precision, and DeepSeek-V4-Pro still requires \SI{795}{\giga\byte}, before accounting for KV (key--value) cache, activations, and
runtime metadata.}}

\gf{On the hardware side, the required storage for weight parameters alone significantly exceeds the on-package memory capacity of recent high-end GPUs~\cite{nvidia2022a100, nvidia2023h100, nvidia2025b200}, which provide only \SIrange{80}{192}{\giga\byte} \gfhpca{of on-package memory capacity, often via multiple \gls{HBM}~\cite{kim2014hbm, jedec2022hbm3}} stacks per \gfhpca{GPU package}~\cite{nvidia2022a100, nvidia2023h100, nvidia2025b200}.
Consequently, deploying frontier-scale \glspl{LLM} often requires partitioning a model across multiple accelerators~\cite{shoeybi2019megatron,huang2019gpipe,fedus2022switch,lepikhin2020gshard,pope2023efficiently}, increasing deployment cost \gfhpca{and introducing interconnect-induced bottlenecks when \acy{partial states} must be communicated across devices~\cite{aminabadi2022deepspeed,yu2022orca,pope2023efficiently}.}}
\gf{
In large-batch datacenter serving, \gfhpca{the workload exploits the arithmetic throughput and memory bandwidth that the added accelerator packages supply for increased performance, and together they hold the model's weights} through tensor, pipeline, or expert parallelism~\cite{aminabadi2022deepspeed,shoeybi2019megatron,yu2022orca}.
\gfhpca{However, in small-batch} single-accelerator and small-node inference settings, such as edge servers, workstations, laptops, and mobile-class systems, adding many accelerator packages solely for model capacity is \gfhpca{often physically and commercially} impractical~\cite{alizadeh2024llmflash,xue2024powerinfer2,sheng2023flexgen}\gfhpca{, even when a single accelerator package could potentially provide enough arithmetic and memory throughput for deployment}.
\revdelhpca{\gfhpca{In our experiments (\cref{sec:eval}), we observe that a \emph{single} accelerator package with enough memory capacity can meet the per-token latency interactive serving currently targets (one output token every \SIrange{40}{100}{\milli\second}~\cite{reddi2020mlperf,
zhong2024distserve,agrawal2024sarathi,mlcommons2025inferencev5}), delivering higher tokens-per-dollar than a capacity-forced multi-accelerator deployment.}}}

\gf{To address the memory-capacity wall, memory vendors~\cite{sandisk2025hbf,ha2026h3} and recent academic proposals~\gfhpca{\cite{suh2026hbfworkload, kim2026hbfroadmap,hsu2026haven,sun2025lincoln,wu2026memexplorer}} have explored emerging memory technologies
as a near-accelerator capacity tier.
A prominent example is \gls{HBF}~\cite{sandisk2025hbf,ha2026h3}, which stacks dense 3D NAND flash dies in an \gls{HBM}~\cite{kim2014hbm, jedec2022hbm3,lee2016simultaneous}-form-factor to provide multi-terabyte capacity per stack with TB/s-class memory bandwidth.
A common approach to integrating \gls{HBF} into the accelerator memory subsystem uses \gls{D2D} links~\cite{sharma2024universal} to daisy-chain \gls{HBF} with \gls{HBM}, which allows expanding on-package capacity while preserving accelerator pin count and simplifying memory controller integration~\cite{ha2026h3,suh2026hbfworkload, kim2026hbfroadmap}.
In this organization, \gls{LLM} weights are preloaded into \gls{HBF} before inference, allowing \gls{HBM} to remain available for latency-critical and dynamically generated state, 
such as the KV cache and intermediate activations \gfhpca{that otherwise could incur high write latency overheads and endurance concerns to the \gls{HBF} memory tier}~\cite{son2026exploring}.
Thus, \gls{HBF} effectively serves as a \emph{read-only}, on-package memory capacity tier for model parameters, making it especially compelling for reducing capacity-driven accelerator count in
single-accelerator and small-node inference systems.
Since \gls{HBF} is built from NAND flash rather than DRAM, its read latency is expected to be microsecond-scale~\cite{ha2026h3,suh2026hbfworkload, kim2026hbfroadmap}, roughly one to two orders of magnitude higher than \gls{HBM}~\cite{jedec2022hbm3}.
To hide this latency \gfhpca{and allow \gls{HBF} to reach read bandwidth on par with \gls{HBM}}, prior designs~\cite{ha2026h3,suh2026hbfworkload,kim2026hbfroadmap}
exploit \emph{inter-layer weight prefetching}~\cite{yuzuguler2025preserve} \gfhpca{alongside \emph{plane-level parallelism}~\cite{abdurrab2013dloop,gao2019parallel,gao2020boosting}}:
while the accelerator computes the current layer using weight blocks already placed in an SRAM staging buffer in the \gls{HBM} base die, compiler- or programmer-directed \gfhpca{prefetching commands move \emph{coarse-grained} weight blocks belonging to the next layer from \gls{HBF} into the SRAM staging buffer} \gfhpca{via a single coordinated multiple-plane flash read operation that collectively returns megabytes' worth of weight data (i.e., a \emph{burst})}. 
Subsequent accelerator requests for model weights\gfhpca{, which are forwarded to \gls{HBF} as \emph{fine-grained} cache misses from the accelerator's \gls{LLC},} can then be served from the staging buffer rather than directly from \gls{HBF}, hiding raw NAND flash access latency. 
In principle, this allows the accelerator to approach \gls{HBM}-side peak memory throughput when the staging buffer is sized
to cover the latency--bandwidth product implied by Little's law~\cite{little1961proof} and when prefetches arrive timely.}

\gf{Although \gls{HBF} offers a promising path to expand near-accelerator memory capacity for \gls{LLM} inference, existing \gls{HBF} proposals~\cite{ha2026h3,suh2026hbfworkload,kim2026hbfroadmap} either oversimplify or leave unresolved several challenges\revdelhpca{ that must be addressed before \gls{HBF} can be efficiently adopted in \gls{LLM} inference systems}.
First, prior designs~\cite{ha2026h3,suh2026hbfworkload,kim2026hbfroadmap} rely on explicit SRAM staging buffers and \gfhpca{static coarse-grained} compiler-emitted prefetch hints to hide \gls{HBF} read latency. 
Our analysis
(\cref{sec:motivation}) shows that this approach leaves two inefficiencies. 
\li~The dedicated SRAM staging buffer introduces avoidable area overhead, because \gls{HBF} \emph{already} contains per-plane page and cache buffers that can collectively stage weight blocks. 
\lii~Static \gfhpca{coarse-grained} compiler-emitted prefetch hints \emph{cannot} reliably align \gls{HBF} burst issue with the \gfhpca{fine-grained cache line} demand stream, since the \gls{HBF} burst consumption order depends on runtime behavior \gfhpca{coming from both the application} (such as \gls{MoE} expert routing, token-dependent attention reuse) \gfhpca{and the hardware} (such as accelerator-side cache line request interleaving), which static layer-ahead prefetch hints \emph{cannot} fully predict.
\revdelhpca{This leaves \gls{HBF} bandwidth unused even when the buffer is sufficiently large.}
Second, prior works~\cite{ha2026h3,suh2026hbfworkload,kim2026hbfroadmap} largely leave refresh management unspecified. Although \gls{HBF} is used as a read-\gfhpca{only weight} tier, 3D NAND \gfhpca{flash} still requires maintenance operations to mitigate retention loss~\cite{luo2018improving,cai2015data,cai2013error,cai2017nand,cai2018reliability} and read-disturb~\cite{luo2018improving,cai2015read,cai2017nand,cai2018reliability,parnell2014modelling}.
Since in \gls{HBF}, a refresh operation can take five orders of magnitude longer than an \gls{HBM} accesses~\cite{cai2017nand,jedec2022hbm3}, scheduling them on the accelerator's critical request path can stall inference for the duration of the operation. 
\revdelhpca{Our analysis shows that, under measured \gls{LLM} read pressure, naive
on-channel refresh would require $10.0\times$--$12.4\times$ the service capacity of a single \gls{HBF} channel before serving any foreground reads.}
Third, \gfhpca{there are missed opportunities to specialize and optimize} the flash-management layer \gfhpca{to the workload access behavior in order to improve its efficiency and reduce its complexity}. 
Conventional \gls{SSD} \gls{FTL}~\cite{cai2017nand,tavakkol2018mqsim,kim2020evanesco,mansouri2022genstore,gal2005algorithms,ftl,intel_ftl,ghiasi2024megis,soysal2025mars,chen2025reis}
is designed for general-purpose storage, where out-of-place updates, garbage collection, and wear leveling are needed to support arbitrary writes. 
However, these \gfhpca{mechanisms} are poorly matched to \gls{LLM} inference, 
\gfhpca{because weights are written once before inference and never updated\revdelhpca{, so \gfhpca{there is \emph{no} need for mechanisms that handle stale data and storage reclaiming since what varies at runtime is the order in which weights are read and \emph{not} which weights are resident in the storage device}.
A \gls{FTL} designed for that behavior can drop the write-management machinery and keep \emph{only} the read path, and to our knowledge no prior \gls{HBF} design specifies one}.}}

\gf{Our \emph{goal} is to design an efficient \gls{HBF} substrate for \gls{LLM} inference that integrates \gls{HBF} as a near-accelerator capacity tier alongside \gls{HBM}, while addressing the adoption challenges identified above.
To this end, we propose \prop (\propLong), a workload-driven \gls{HBF} substrate.
\gfhpca{The \emph{key idea} of \prop is to drive the \gls{HBF} substrate from the \emph{dynamic} read access stream the accelerator actually produces, rather than from a \emph{static} compile-time prediction.
During runtime, the accelerator issues many fine-grained \gls{HBF} read requests that arrive out-of-order but concentrate on a small number of weight blocks.
From that, \prop dynamically reconstructs a coarse-grained burst request by ordering and coalescing fine-grained read requests; thus, multiple fine-grained requests are serviced with \emph{one} coordinated multi-plane NAND page-read latency.
\prop applies the same principle to read-disturb mitigation and to address
translation via three mechanisms.}}

\gf{First, \prop replaces compiler-driven prefetching with a hardware \emph{burst-buffer controller} co-located with the \gls{HBF} base die.
Because \gls{LLM} inference accesses weights through predictable burst-granularity sweeps, the burst-buffer controller serves cache line requests to the active burst and \emph{proactively} issues an \gls{HBF} request to the next burst identified by a lookahead window \emph{before} the active burst finishes draining.
This keeps the \gls{HBF} channel continuously occupied. 
While the current burst streams to the accelerators, \prop has already issued the next burst request to \gls{HBF}, \gfhpca{because it times that request from the progress of the active burst rather than from a program point fixed at compile time.}
\gf{By moving burst timing into hardware, \prop \gfhpca{provides two main benefits. It} \gfhpca{\li~}eliminates compiler-inserted prefetch directives and keeps the \gls{HBF} tier transparent to existing inference kernels and software stacks; and}
\gfhpca{\lii~}removes the need for a dedicated SRAM staging buffer, since the page and cache buffers already present inside each \gls{HBF} die collectively provide enough buffering for two in-flight bursts.
\revdelhpca{Together, these effects amortize one NAND page-read latency across thousands of cache line accesses while avoiding a separate on-package SRAM staging structure.}}

\gf{Second, \prop introduces \emph{phantom-plane refresh}, a continuous refresh mechanism that exploits the \acy{read-only} nature of \gls{LLM} weight access.
Each \gls{HBF} die is provisioned with $N{+}1$ physical planes for an $N$-plane application address space.
At any moment, exactly one of the $N{+}1$ physical planes is offline and plays the role of \emph{phantom plane}, while the remaining $N$ planes serve read requests at full bandwidth.
\revdelhpca{The phantom role is \emph{not} pinned to a single physical plane.
Instead, it rotates round-robin across all $N{+}1$ planes, so every plane spends most of its time in the live set and only briefly assumes the phantom role to be refreshed.}
\gfhpca{When a block on a live plane $A$ reaches its read-disturb threshold, the phantom-plane refresh controller forks the block's contents from the foreground read path and programs an \gls{ECC}-corrected copy into a free block of the phantom plane, using $P$'s independent program circuitry and avoiding contention with the reads on $A$.
\revdelhpca{Once the program completes, the block's entry in a block-granular translation table is updated so that subsequent reads hit the fresh copy, and the vacated block returns to a free-block pool to be erased off the read stripe.}}
Since flash programming operations are only executed in $P$, such operations (which dominate a refresh's cost)
\emph{never} occupy a plane that is serving decode reads.
\revdelhpca{Therefore, refresh becomes a steady-state background operation rather than an accelerator-visible scheduling event, eliminating refresh-induced stalls on the decode critical path.
\gfhpca{\prop costs one additional plane per die, $1/N$ of the area of an $N$-plane die, and preserves advertised capacity and foreground read bandwidth.
The binding cost is \emph{endurance}, \emph{not} area, because every refresh spends a program/erase cycle and read disturbance therefore sets the device's service life.
Rotating relocated blocks through the free-block pool levels wear across physical blocks \emph{without} a dedicated wear-leveling policy, reaching 0.8 to 8.0 years of service life under continuous full-rate decode as \gls{SLC} endurance ranges from $10^{6}$ to a projected $10^{7}$ program/erase cycles.}}}
\gf{Third, \prop adopts a simplified, \emph{read-only \gls{FTL} design} matched to the regularity of \gls{LLM} weight access.
Because each \gls{HBF} access fetches a full burst, address translation is maintained at burst granularity rather than page granularity.
For a \SI{512}{\giga\byte} Gen-1 \gls{HBF} stack with \SI{2}{\mega\byte} bursts, this requires only  256~K entries, \gfhpca{$512\times$ fewer than a conventional page-level \gls{SSD} \gls{FTL}~\cite{cai2017nand,tavakkol2018mqsim}, because one burst entry replaces the 512
page entries it fetches in parallel.}
\revdelhpca{At model load, \prop creates a simple burst-to-plane-set mapping across the address space.
Because \gls{LLM} inference does \emph{not} modify \gls{HBF}-resident weights after the initial model-load phase, \prop removes SSD-style out-of-place updates and garbage collection from the foreground access path, while wear is managed by phantom-plane rotation.}
\gfhpca{The resulting \gls{FTL} reduces to a burst translation table, a block-granular relocation table, and one read counter per block, together 1.8\,MB per stack, which replaces the large, latency-sensitive metadata path of an \gls{SSD}-class \gls{FTL} with a compact, constant-time structure that fits in the \gls{HBF} base die.}}

\gf{We evaluate \prop on \gfhpca{six production \glspl{LLM}: five \gls{MoE} models (DeepSeek-V3~\cite{liu2024deepseek}, DeepSeek-V4-Pro~\cite{deepseekai2026deepseekv4}, Qwen3-235B-A22B~\cite{bai2025qwen3}, Llama-4 Maverick~\cite{meta2025llama}, and Kimi~K2~\cite{team2025kimi}) and one dense model (Llama-3.1-405B~\cite{dubey2024llama3})}.
We compare \prop against three baselines: 
\li~\emph{HBM+SSD}, a single-GPU system that spills model weights to fast off-package \glspl{SSD} once they exceed \gls{HBM} capacity; 
\lii~\emph{HBM-only}, a multi-GPU system that shards the model across enough \gfhpca{GPU packages} to keep all weights in aggregate \gls{HBM}; and
\liii~\emph{H$^3$}~\cite{ha2026h3}, a prior \gls{HBM}+\gls{HBF} design that relies on an \gls{HBM}-side staging buffer and compiler-emitted layer-ahead prefetch hints.
Our results show that \prop
\li~sustains $6.2\times$ higher effective \gls{HBF} throughput than prior \gls{HBF} designs;
\lii~improves decode throughput by $1{,}205\times$, $2.2\times$, and
$6.2\times$, and reduces energy consumption by $408\times$, $1.1\times$, and $6.8\times$ that of an \gls{SSD}-equipped GPU system, an \gls{HBM}-only GPU system, and a prior hybrid \gls{HBM}+\gls{HBF} system, respectively; 
\liii~meets a \SI{50}{\milli\second} \gls{TPOT} \gls{SLO} with $3.1\times$ fewer GPU packages than the \gls{HBM}-only GPU system (up to $8\times$); and
\liv~incurs small area cost to an \gls{HBF} die ($3.1~\%$) and
\gls{HBF} base die (\SI{3.9}{\milli\meter\squared} at
\SI{7}{\nano\meter}).}

\gf{We make the following key contributions:}
\begin{itemize}[noitemsep,topsep=0pt,parsep=0pt,partopsep=0pt,%
                labelindent=0pt,itemindent=0pt,leftmargin=*]
    \item \gf{We identify three key adoption challenges for \gls{HBF}-tiered \gls{LLM} inference: \gfhpca{coarse-grained} static prefetching underutilizes the \gls{HBF} channel, on-channel refresh oversubscribes the foreground read path, and \gls{SSD}-class \glspl{FTL} overprovision metadata and write-path hardware for read-only weights.}

    \item \gf{We propose \prop, a workload-driven \gls{HBF} substrate that co-designs three mechanisms around the regularity of \gls{LLM} weight access: \li~a \emph{hardware burst-buffer controller} that uses the page and cache buffers already inside each \gls{HBF} die instead of a dedicated SRAM staging buffer, \lii~\emph{phantom-plane refresh} that keeps refresh off the decode critical path, and \liii~a \emph{read-only \gls{FTL}}, which replaces \gls{SSD}-class support for arbitrary writes with compact logical-to-physical address translation for \gls{HBF}-resident weight bursts.}

    \item \gf{We evaluate \prop on dense and \gls{MoE} models, showing that \prop increases decode throughput and reduces energy consumption.}
\end{itemize}
 
\section{Background}
\label{sec:background}

\begin{figure}[ht]
    \centering
    \includegraphics[width=\linewidth]{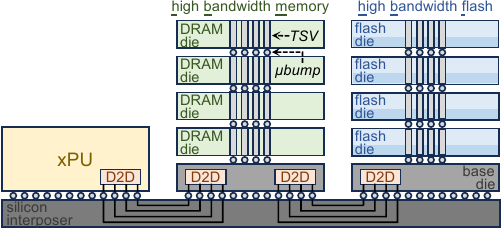}
    \caption{\gf{Overview of an \gls{HBF}-enabled xPU package.}}
    \label{fig:hbfoverview}
\end{figure}

\paratitle{\gf{\gls{HBF}-Enabled System}}
\gf{Recent \gls{HBF} proposals~\cite{ha2026h3,suh2026hbfworkload,kim2026hbfroadmap,sandisk2025hbf} integrate NAND flash memory~\cite{bez2003introduction} directly into the accelerator memory subsystem. 
Fig.~\ref{fig:hbfoverview} shows a representative \gls{HBF}-enabled package, where xPUs\revdelhpca{(e.g., \glspl{GPU}~\cite{nvidia1999transformlighting,nvidia2022a100,nvidia2023h100,nvidia2025b200}, \glspl{TPU}~\cite{jouppi2017datacenter}, \glspl{NPU}~\cite{chen2014diannao,liao2021ascend})}, \gls{HBM} stacks~\cite{jedec2022hbm3,kim2014hbm,lee2016simultaneous}, and \gls{HBF} stacks~\cite{ha2026h3,suh2026hbfworkload,kim2026hbfroadmap,sandisk2025hbf} are co-packaged on a silicon interposer. 
Each \gls{HBM} stack vertically bonds 12--16 DRAM dies over a logic-bearing \emph{base die}, providing \SIrange{24}{36}{GB} of capacity and \SIrange{1.2}{2}{\tera\byte\per\second} of \gfhpca{memory} bandwidth per stack~\cite{jedec2022hbm3}. 
\gls{HBF} follows a similar form factor, stacking NAND flash dies (e.g., 16) over an \gls{HBF} base die to provide \SIrange{512}{1024}{GB} of capacity and \SIrange{1.6}{2}{\tera\byte\per\second} of aggregate NAND bandwidth per stack~\cite{sandisk2025hbf,ha2026h3}. 
In both memories, thousands of \glspl{TSV} and \(\mu\)bumps connect the memory dies to the base die.}
\gf{The base dies implement the control and interconnect logic needed to compose these two memories. 
The \gls{HBM} base die contains DRAM channel controllers, refresh and \gls{ECC} logic, HBM PHYs, and a \gls{D2D} PHY~\cite{sharma2024universal} toward the xPU. 
\gfhpca{Following current \gls{HBF} literature~\cite{suh2026hbfworkload,ha2026h3,son2026exploring}, the} \gls{HBF} base die contains the flash controller, \gfhpca{which we model as comprising} the \gls{FTL}~\cite{cai2017nand,tavakkol2018mqsim}, NAND command sequencer, NAND \gls{ECC} engine, and a \gls{D2D} PHY toward the \gls{HBM} base die. 
Thus, the xPU connects to \gls{HBM} through one \gls{D2D} link~\cite{sharma2024universal}, while \gls{HBM} connects to \gls{HBF} through another, daisy-chaining both tiers behind a single xPU-facing shoreline budget~\cite{ha2026h3}.}

\gf{This physical integration increases memory capacity in the \gfhpca{xPU package}, but making flash usable as an xPU memory tier also requires hiding NAND latency and preserving a familiar memory interface. 
Therefore, current proposals~\cite{ha2026h3,suh2026hbfworkload,kim2026hbfroadmap} expose \gls{HBM} and \gls{HBF} as a unified physical address space, with \gls{HBF} accessed transparently through the \gls{HBM} base die~\cite{ha2026h3,suh2026hbfworkload,kim2026hbfroadmap}. 
To support this model, the \gls{HBM} base die adds three mechanisms. 
First, an \emph{address decoder} partitions the address space into \li~an \gls{HBM}-resident region for latency-sensitive \gfhpca{and write-mostly} 
state, such as KV cache, activations, and runtime metadata, and 
\lii~an \gls{HBF}-resident region for read-\gfhpca{only} inference state, such as \gls{LLM} weights preloaded into \gls{HBF} at deployment time. 
The \gfhpca{address} decoder routes each xPU request \gfhpca{(in the form of \emph{fine-grained} \gls{LLC} misses in a \gls{GPU}-based system)} either to local \gls{HBM} banks or through the \gls{HBM}$\rightarrow$\gls{HBF} \gls{D2D} link, to the \gls{HBF} stack.
Second, a \emph{latency-hiding buffer} (LHB)~\cite{ha2026h3}, implemented as megabyte-scale SRAM in the \gls{HBM} base die, stages prefetched weight blocks from \gls{HBF}. 
This buffer hides NAND page-read latency, which is in the microsecond range~\cite{ha2026h3,suh2026hbfworkload} and two to three orders of magnitude longer than an \gls{HBM} access~\cite{jedec2022hbm3}. 
Third, a \emph{prefetch controller} consumes compiler- or programmer-emitted prefetch directives~\cite{yuzuguler2025preserve} and refills the LHB over the \gls{HBM}$\rightarrow$\gls{HBF} \gls{D2D} link.
\revdelhpca{In steady state, xPU requests to the HBF region hit in the LHB and observe SRAM-class latency, while NAND accesses are overlapped with computation.}
}

\begin{figure}[ht]
    \centering
    \includegraphics[width=\linewidth]{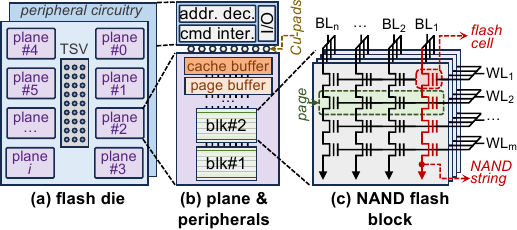}
    \caption{\gf{Overview of a NAND flash die organization.}}
    \label{fig:flash-organization}
\end{figure}

\paratitle{\gf{\gls{HBF} Organization}} \gf{A 3D NAND flash subsystem is organized as a hierarchy of components, as Fig.~\ref{fig:flash-organization} illustrates. 
An \gls{HBF} stack (Fig.~\ref{fig:hbfoverview}) contains multiple (e.g., 16) flash dies~\cite{sandisk2025hbf,ha2026h3}. 
Each \emph{flash die} (Fig.~\ref{fig:flash-organization}a) integrates tens (e.g., 32) of flash \emph{planes} around a central \gls{TSV} array that propagates command, address, and data signals through the die stack~\cite{kim2026hbfroadmap,suh2026hbfworkload}. 
\revdelhpca{The flash array is bonded face-to-face with a \emph{peripheral CMOS} layer that hosts the per-die address decoder, command interface, and I/O drivers; the two layers are joined using Cu-pad hybrid bonds, following the \gls{CBA} scheme adopted by recent 3D NAND generations~\cite{goda2021nand3d,sako20231tb,kobayashi2023high}.}
A flash \emph{plane} (Fig.~\ref{fig:flash-organization}b) consists of hundreds (e.g., 256) of NAND flash \emph{blocks} and two small SRAM structures in the peripheral CMOS layer: the \emph{page buffer} and the \emph{cache buffer}~\cite{cai2017nand}. 
The page buffer latches the page currently being sensed or programmed, while the cache buffer holds the previous sensed page\revdelhpca{ so that draining data to the host can overlap with the next NAND sense, improving streaming-read throughput}.}
\gf{A NAND flash \emph{block} (Fig.~\ref{fig:flash-organization}c) is the smallest erase unit and is organized as a 3D array of \emph{NAND strings}. 
A NAND string is a vertical column of hundreds (e.g., 256) of \emph{flash cells} along a bitline. 
The string connects through a string-select transistor to a bitline (BL) at the top and to a common source line at the bottom. 
Multiple (e.g., 4) NAND strings share each bitline through their respective string-select transistors, allowing each block to contain thousands of pages (e.g., 1024 pages). 
Flash cells in the same horizontal layer share a \emph{wordline} (WL); together with the intersecting bitlines and the selected
string, a wordline forms one \emph{page} (Fig.~\ref{fig:flash-organization}, highlighted), the smallest read and
program unit.
With this organization, an \gls{HBF} stack with 16 dies, 32 planes per die, 256--512 blocks per plane, 1024 pages per block, and \SI{4}{\kilo\byte} pages provides \SIrange{512}{1024}{GB} of \gls{SLC} (i.e., one bit per flash cell with two threshold-voltage levels) capacity per module, spanning current \gls{HBF} Gen-1 and upcoming Gen-2 design points~\cite{sandisk2025hbf,kim2026hbfroadmap}.}

\paratitle{\gf{NAND Flash Operation}} \gf{A flash cell stores one or more bits as charge trapped in the charge-trap layer surrounding its channel. 
NAND flash supports three primitive operations: \emph{page read}, \emph{page program}, and \emph{block erase}~\cite{cai2017nand}. 
First, a \emph{page read} senses the threshold voltage of every flash cell on a selected wordline and transfers the data to the page buffer. 
Operating flash cells as \gls{SLC} requires a single sense operation per read and yields a page-read latency $t_R$ of \SIrange{1}{2}{\micro\second} in \emph{modern} \gls{SLC} NAND~\gfhpca{\cite{suh2026hbfworkload,sandisk2025hbf, shiozawa2020emerging,cheong2018flash}}.
Second, a \emph{page program} injects charge into the cells of a selected wordline, transitioning them from the erased state to a programmed state. 
A programmed page \emph{cannot} be overwritten without first erasing its block.
A page program takes approximately \SI{50}{\micro\second}.
Third, a \emph{block erase} removes trapped charge from every cell in a block, returning the block to the erased state, and is the only way to clear programmed cells. 
A block erase takes approximately \SI{3}{\milli\second}~\cite{cai2017nand}.}

\paratitle{\gfhpca{Read Disturbance in Flash}} \gf{Repeated operations and time gradually shift flash cell threshold voltages, eventually causing read errors~\cite{luo2018improving}. 
Two mechanisms dominate this drift. 
First, \emph{read-disturb} accumulates over many reads to the same block, because each read biases all non-selected wordlines to a high pass-through voltage~\cite{cai2015read}. 
Second, \emph{retention loss} accumulates over time as trapped charge slowly leaks from the cells~\cite{cai2015data}.
To bound the resulting bit-error rate, the flash controller periodically performs \emph{refresh}: it reads a block, ECC-corrects any errors, reprograms the corrected data to a fresh location, and erases the stale block. 
Refreshing one \gls{SLC} block costs one block erase, approximately \SI{3}{\milli\second}, plus one program per page, e.g., 1024 pages at approximately \SI{50}{\micro\second} each, for a total of about \SI{55}{\milli\second}~\cite{cai2017nand,luo2018improving}. 
For \gls{SLC} NAND, a block must be refreshed after roughly $10^5$--$10^6$ reads to that block due to read-disturb, or after several years due to retention loss~\cite{cai2017nand,cai2018reliability}.}  

\section{\gf{Motivation}}
\label{sec:motivation}

\gf{\gfhpca{In this section, we first highlight the opportunities that an \gls{HBF}-equipped system provides for capacity-bound inference regimes. Second, we} identify three major shortcomings of prior \gls{HBF}
architectures for \gls{LLM} inference:
\li~static prefetching underutilizes the \gls{HBF} channel;
\lii~refresh competes with foreground reads and can saturate the \gls{HBF} channel; and
\liii~\gls{SSD}-class \glspl{FTL} retain unnecessary write-path machinery and overprovisioned metadata for a read-mostly inference workload.}

\subsection{\gfhpca{The Case for \gls{HBF} as a Capacity Weight Tier}}
\label{sec:motiv:case}

\revdelhpca{\gfhpca{Serving a frontier-scale \gls{LLM} starts from a storage capacity constraint. 
\revdelhpca{At their native data precision, the weights of DeepSeek-V3 occupy \SI{1.3}{\tera\byte}, Kimi~K2 \SI{1.0}{\tera\byte}, Llama~3.1-405B \SI{810}{\giga\byte}, and DeepSeek-V4-Pro
\SI{795}{\giga\byte}, while one high-end GPU package provides around \SI{192}{\giga\byte} of \gls{HBM}~\cite{nvidia2025b200} for storage.
Thus, deploying such models would require sharding the model
across at least four to six \gls{GPU} packages and coupling them with tensor, pipeline, or expert parallelism~\cite{shoeybi2019megatron,huang2019gpipe,lepikhin2020gshard,aminabadi2022deepspeed}.}
However, we observe that such \emph{capacity-driven} deployment leads to several inefficiencies depending on the target batch size for the system.}}

\begin{figure}[ht]
    \centering
    \includegraphics[width=0.95\linewidth]{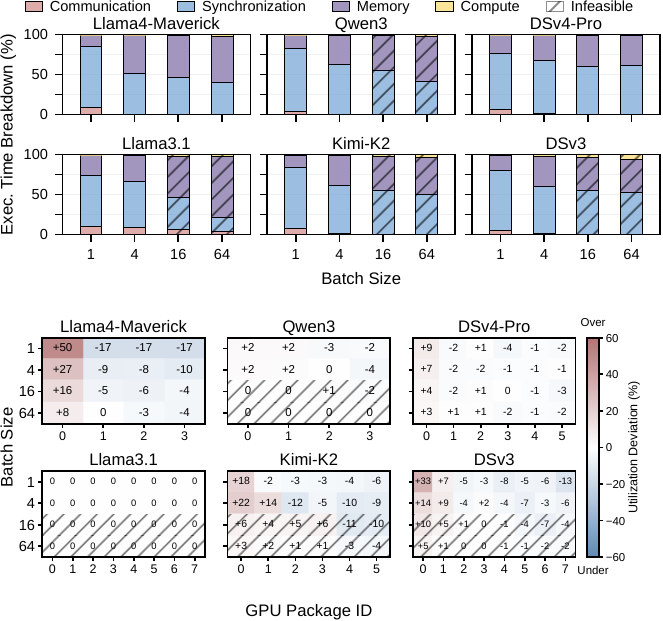}
    \caption{\gfhpca{Cost of a minimum-fit \gls{HBM}-only deployment at 128K~tokens of context for various batch sizes: (top)~execution time breakdown, (bottom)~memory traffic imbalance.}}
      \label{fig:motiv-scaleout}
\end{figure}

\paratitle{\gfhpca{Limitations of Capacity-Forced Scaled-Out Inference}} \gfhpca{We observe that a capacity-driven \gls{LLM} inference deployment leads to communication, synchronization, and load imbalance overheads. 
Fig.~\ref{fig:motiv-scaleout} (top) decomposes per-token time for the \emph{minimal-fit deployment} (i.e., the minimal number of GPU packages required to store the model weights in their target data precision), with a context size of 128K tokens and varying batch sizes ($bsz \in \{1,4,16,64\}$) for six different \glspl{LLM} (i.e., 
Llama-4 Maverick~\cite{meta2025llama} with four B200 \gls{GPU} packages, 
Qwen3-235B-A22B~\cite{bai2025qwen3} with four B200 \gls{GPU} packages,
DeepSeek-V4-Pro~\cite{deepseekai2026deepseekv4} with six B200 \gls{GPU} packages,
Llama-3.1-405B~\cite{dubey2024llama3} with eight B200 \gls{GPU} packages,
Kimi~K2~\cite{team2025kimi} with six B200 \gls{GPU} packages, and
DeepSeek-V3~\cite{liu2024deepseek} with eight B200 \gls{GPU} packages); see \cref{sec:methodology} for our experimental methodology. 
We split the per-token time into communication, synchronization, memory, and compute. 
Fig.~\ref{fig:motiv-scaleout} (bottom) shows how evenly the memory traffic of a decode step is spread over the GPU packages: each cell reports how far, in percent, one package's memory-bandwidth utilization is above or below the average memory traffic. 
\revdelhpca{Zero means the package moves exactly as many bytes as the average package.}
We make three observations.}

\gfhpca{First, inter-package communication and synchronization together account for \SI{52}{\percent} to \SI{85}{\percent} of per-token time at small batch sizes ($bsz\!=\!1$ to $4$), and still \SI{22}{\percent} to \SI{61}{\percent} by $bsz\!=\!64$.
\revdelhpca{In the five expert-parallel models, each \gls{MoE} layer transfers data twice: 
\li~before the experts run, every package sends each of its tokens to whichever package holds the expert the router selected for that token; 
\lii~after the experts run, every package sends each result back to the package the token came from. 
Both transfers are \emph{all-to-all}, because a package holds tokens bound for experts on every other package, and attention adds its own transfers across the packages that shard it. 
In tensor-parallel Llama3.1, each layer instead sums partial results across all GPU packages before the next layer reads them. 
A GPU package \emph{cannot} proceed past an exchange until the data it is owed arrives, so a token pays the last arrival rather than the average, and the number of such waits a token pays is set by the model's depth rather than by batch size.}
Second, a significant portion of the execution time is spent waiting at the layer barrier (\SI{51}{\percent} to \SI{79}{\percent} of per-token time at $bsz\!=\!1$ to $4$) since the physical inter-package communication is rarely blocking (only \SI{0.4}{\percent} to \SI{9.7}{\percent} of inter-package communication blocks computation).
The primary reason for the high synchronization overhead is \emph{expert routing imbalance}.
The busiest GPU package reads 1.4$\times$ to 4$\times$ the routed-expert weight bytes of the mean package over that range, falling to 1.1$\times$ by $bsz\!=\!64$, because one token routes to at most 8 of a model's 128--384 experts and therefore \emph{cannot} load every GPU package evenly, while a larger batch routes enough tokens to reach and fill all of them.
The GPU packages that finish early hold at the barrier until the busiest one
arrives. 
Third, accelerator compute occupies \emph{only} \SI{0.1}{\percent} to \SI{1.9}{\percent} of per-token time at $bsz\!=\!1$ to $4$), rising to \SI{5.9}{\percent} by $bsz\!=\!64$\revdelhpca{, since batching adds arithmetic to a decode step without adding transfers or barriers to it}.}
\revdelhpca{\gfhpca{We conclude that a minimum-fit deployment spends \emph{significant} per-token time on communication and synchronization at small batch sizes, and such overheads are amortized only once the batch is large enough.
We call the small-batch regime \emph{capacity-forced scale-out}, since
the accelerators are provisioned to hold weights rather than to supply
arithmetic.}}

\begin{figure}[ht]
  \centering
  \includegraphics[width=0.9\linewidth]{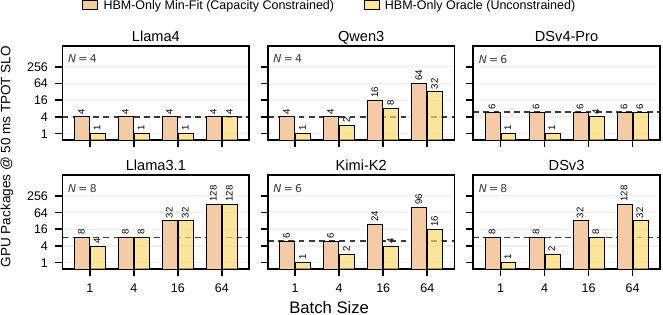}
  \caption{\gfhpca{GPU packages needed to hold the model and meet the
    \SI{50}{\milli\second} TPOT \gls{SLO} at a context size of 128K~tokens\revdelhpca{, for the minimum-fit \gls{HBM}-only deployment and the same system with its capacity constraint lifted}.}}
  \label{fig:motiv-oracle}
\end{figure}

\paratitle{\gfhpca{Lifting Capacity-Forced Scale-Out}}
\gfhpca{To isolate what the capacity constraint alone costs, we compare the minimum-fit deployment against a \emph{capacity-lifted} oracle, the same \gls{HBM}-only system with the weight-capacity constraint waived, so that weights and KV cache always fit within a \emph{single} \gls{GPU} package. 
In this model, \gls{GPU} packages are replicated \emph{only} for the sake for higher memory and compute throughput. Fig.~\ref{fig:motiv-oracle} reports the
GPU packages each system needs to serve different \gls{LLM} inference models within a \SI{50}{\milli\second} \gls{TPOT} \gls{SLO}. 
We make three observations.
First, at $bsz\!=\!1$ the oracle serves every \gls{MoE} model within the \gls{SLO} on a \emph{single} \gls{GPU} package, while the minimum-fit deployment holds four to eight, so the capacity constraint overprovisions \gls{GPU} packages by $4.6\times$ on average.
Second, the overprovisioning shrinks as the batch grows but does \emph{not}
vanish ($1.9\times$ overprovisioning for $bsz\!=\!64$).
Third, dense Llama~3.1-405B provides a different trend, only saving
$2\times$ \gls{GPU} packages at $bsz\!=\!1$ and providing \emph{no} savings at $bsz\!=\!64$ (128 packages in both systems), because reading every weight once per token demands the aggregate bandwidth of many packages.
\revdelhpca{We conclude that the capacity constraint alone costs the minimum-fit deployment $2.8\times$ its package bill on average across all models and batch sizes at 128K~tokens of context, and that the cost concentrates in the small-batch regime.}}
\revdelhpca{\gf{Unfortunately, \gls{HBM} \emph{cannot} supply the missing capacity within a \gls{GPU} package's \gls{HBM} stack budget, since \gls{HBM} capacity per stack has grown far more slowly than parameter counts~\cite{kim2014hbm,jedec2022hbm3,gholami2024airadar}. 
\gls{HBF} supplies it instead by stacking dense 3D NAND flash dies in an \gls{HBM} form factor\revdelhpca{ behind the same package boundary~\cite{sandisk2025hbf,ha2026h3}, offering multi-terabyte capacity per stack at \unit{TB/s}-class read bandwidth~\cite{ha2026h3,suh2026hbfworkload,kim2026hbfroadmap}}. 
What \gls{HBF} does \emph{not} offer is \gls{HBM}-class read latency, since a NAND flash page sense takes one to two orders of magnitude longer than an \gls{HBM} access~\cite{cai2017nand,jedec2022hbm3}.}\revdelhpca{We conclude that an on-package flash tier removes the capacity constraint that forces scale-out at small batch, and that whether an \gls{HBF}-tiered
accelerator matches the package bill of the capacity-lifted oracle in
Fig.~\ref{fig:motiv-oracle} is decided by how efficiently the substrate can hide flash read latency behind the decode stream.}}

\subsection{\gf{Limitations of State-of-the-Art \gls{HBF}-Based Systems}}

\begin{figure}[ht]
  \centering
  \includegraphics[width=\linewidth]{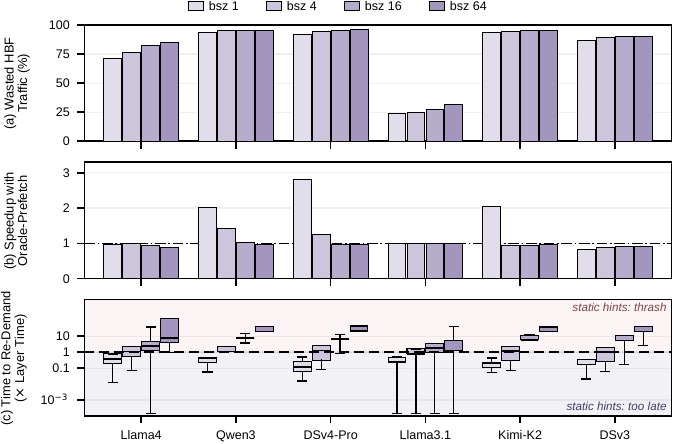}  \caption{\gfhpca{\gls{HBF} channel usage under the static H$^{3}$ prefetcher.
  \revdelhpca{at a 128K tokens context size. 
  (a)~Wasted share of H$^{3}$'s burst traffic. 
  (b)~Speedup of an oracle prefetcher with perfect hint coverage over H$^{3}$. 
  (c)~Time from a burst's eviction to its next demand, in multiples of one layer's execution time (boxes span the 25th to 75th percentiles, whiskers the 5th to 95th).}}}
  \label{fig:motiv-h3}
\end{figure}

\paratitle{\gf{Static Prefetch Underutilizes the \gls{HBF} Channel}}
\gf{An \gls{HBF} read operation begins with a NAND flash page sensing, which takes $t_R$~ns (e.g., \SIrange{1}{2}{\micro\second}) and returns a few kilobytes of data (e.g., a \SI{4}{\kilo\byte} page) per plane. 
To approach the \gls{HBM} channel's \si{\tera\byte\per\second}-class bandwidth, the \gls{HBF} controller must amortize this fixed page read latency across many planes:
for example, a coordinated sense at the same \texttt{(block, page)} coordinate across all 512 planes of an \gls{HBF} stack returns one
\SI{2}{\mega\byte} burst in one $t_R$. 
This burst size matches the channel bandwidth-delay product, $\text{BDP}=\SI{1}{\tera\byte\per\second}\times t_R
=\SI{2}{\mega\byte}$, and is therefore the \emph{minimum} granularity needed to sustain peak read \gls{HBF} bandwidth. 
Prior \gls{HBF}
designs~\cite{ha2026h3,suh2026hbfworkload,kim2026hbfroadmap} adopt this BDP-sized operating point by provisioning two staging-buffer slots: one burst drains to the GPU while the next burst is in flight from \gls{HBF}. 
These slots are filled using compiler-emitted,
per-layer prefetch hints for model weights~\cite{yuzuguler2025preserve}, with the goal of hiding NAND flash page-read latency from the \gls{GPU}.}
\gf{\revdelhpca{At this sizing, achieving \gls{HBF} peak bandwidth depends not only on \emph{what} is prefetched but also on \emph{when} each burst request is issued, since the next burst must arrive in the staging buffer by the time the current burst finishes draining.}
\revdelhpca{However, a two-slot BDP-sized buffer provides little slack: if the correct burst is \emph{not} already in flight when the \gls{SM} demand stream reaches it, the GPU is exposed to the full NAND flash page-read latency.
Compiler-driven prefetching must choose the burst-issue order \emph{prior} to execution, but the actual consumption order is shaped by runtime behavior that the compiler cannot fully know, including \gls{MoE} expert routing, token-dependent attention reuse, and \gls{SM}-level interleaving.}}
\gfhpca{To measure how a static controller uses the \gls{HBF} channel, we run the prior-art H$^{3}$ design~\cite{ha2026h3}, which prefetches bursts one layer ahead in compile-time order, and classify every fetched byte from \gls{HBF} as consumed by the GPU or wasted\revdelhpca{, where a wasted byte leaves the staging buffer before any \gls{SM} reads it}. 
Fig.~\ref{fig:motiv-h3}(a) shows the wasted fraction of H$^{3}$'s burst traffic at a context size of 128K tokens, across $bsz\in\{1,4,16,64\}$, and panels (b) and (c) trace the waste to its root cause. 
We make three observations.}

\gfhpca{First, the waste is large for every architecture. The \gls{MoE} models waste 86~\% to 96~\% of the traffic at every batch size.
Second, the waste is \emph{not} a prefetch coverage problem. 
Fig.~\ref{fig:motiv-h3}(b) grants
H$^{3}$ an oracle prefetcher, a hint for \emph{every} candidate burst issued at the same layer-ahead timing. 
Perfect coverage changes serving-batch throughput by \emph{at most} 3~\% ($0.90\times$ to $1.03\times$ at $bsz\!=\!16$ to $64$ across all six models), recovers $2.0\times$ to $2.8\times$ on three \gls{MoE} models only at batch~1, and strictly \emph{hurts} DeepSeek-V3 at every batch ($0.83\times$ to $0.92\times$), because the full-coverage stream evicts live bursts to make room for prefetched ones.
Third, we observe that a re-fetch either happens due to prefetch hints being too late or due to thrashing of the data in the staging buffer, as Fig.~\ref{fig:motiv-h3}(c) shows. 
In the figure, for every re-fetched burst, we measure the time from its eviction to its next demand, as a multiple of one layer's execution time; a value of 1 means the burst was
demanded again exactly one layer's time after leaving the buffer, which is also the maximum head start a layer-ahead hint can have, so the dashed line at 1 marks the \emph{earliest} any static hint can act. 
Re-fetches in the lower region return sooner than that, so a hint arrives structurally \emph{too late}. 
We note that at  $bsz\!=\!1$, the entire distribution sits in the \emph{too late} region in every model, with median return times of $0.12\times$ to $0.45\times$ the layer time. 
Re-fetches in the upper region, which dominate at $bsz\!=\!64$, leave time for a prefetch hint to act, but the per-layer active set then exceeds the staging buffer capacity, so the re-issued prefetch \emph{thrashes} the buffer by evicting other live bursts.
\revdelhpca{We conclude that H$^{3}$'s waste is set by \emph{when} bursts are issued and what the staging buffer retains, \emph{not} by which addresses the compiler knows.}}

\begin{figure}[ht]
  \centering
  \includegraphics[width=\linewidth]{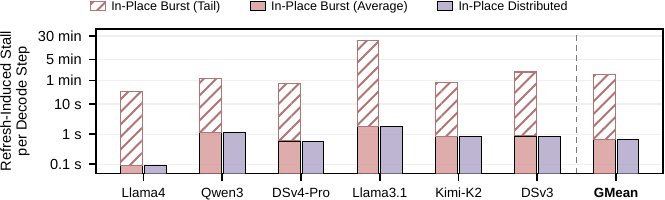}
  \caption{\gfhpca{Refresh-induced stall per decode step\revdelhpca{ of in-place refresh under the H$^{3}$ baseline}.}}
  \label{fig:motiv-refresh}
\end{figure}

\paratitle{\gf{On-Channel Refresh Saturates the \gls{HBF} Channel}}
\gf{A na\"ive \gls{HBF} controller would refresh blocks on the same channel used by foreground reads, which is prohibitively expensive.
A single block refresh costs one block erase plus one program per page:
$t_\text{refresh}=t_\text{ERASE}+1024\times\,t_\text{PROG} =\SI{54.2}{\milli\second}$ at $t_\text{ERASE}=\SI{3}{\milli\second}$ and $t_\text{PROG}=\SI{50}{\micro\second}$. 
\gfhpca{Reading the block's 1024~pages out before the erase adds up to $1024\times t_\text{R}$ on top, which we conservatively omit.} 
This is more than four orders of magnitude longer than a single page read and stalls \emph{all} foreground reads to the refreshed \gfhpca{block's plane} for the full duration.}
\gfhpca{We quantify this refresh pressure under the H$^{3}$ baseline. Each flash block tolerates a conservative \gls{SLC} read-disturb threshold of $N_\text{disturb}=10^{6}$ page reads~\cite{ha2026h3,cai2017nand} before it must refresh. 
Intuitively, refresh operations can be triggered by at least two mechanisms:
\li~\emph{in-place burst refresh}, where all per-block read counters are initialized together, and all always-on blocks reach their refresh threshold on the same token; 
\lii~\emph{in-place distributed refresh}, where per-block read counters are initialized spread apart, allowing the steady refresh stream to be interleaved with foreground reads. Fig.~\ref{fig:motiv-refresh} shows the resulting refresh-induced stall per decode step for the six \glspl{LLM} at $bsz\!=\!1$ and a context length of 128K tokens. 
We make two observations.}
\gfhpca{First, in-place burst refresh stalls the \gls{HBF} channel for \SI{26}{\second} to \SI{22}{\minute} whenever the accumulated refreshes are due, because the aligned read counters bring thousands of blocks due simultaneously, and each block refresh operation occupies the channel for \SI{54.2}{\milli\second}.
Second, in-place distributed refresh improves over burst by spreading the refresh work over the time dimension.
However, the steady refresh stream still inserts \SI{0.7}{\second} of refresh service into every decode step on average (\SI{0.09}{\second} to \SI{1.8}{\second}).
\revdelhpca{We conclude that both in-place refresh mechanisms impose non-trivial refresh stall directly on the decode execution path.}}

\paratitle{\gf{\gls{SSD}-Class \gls{FTL} Overprovisions the Read Path}}
\gf{A conventional \gls{SSD} \gls{FTL}~\cite{cai2017nand,tavakkol2018mqsim} is designed for arbitrary host writes. 
Therefore, it includes hardware units for out-of-place updates, garbage collection, wear leveling, free-block allocation, and crash consistency. 
However, these mechanisms are \emph{unnecessary} for an \gls{HBF} serving as a weight tier for \gls{LLM} inference, where model weights are programmed once at deployment time and remain read-only during inference.
Table~\ref{tab:ftl-components} summarizes the major \gls{SSD}-class \gls{FTL} components, their approximate cost at the \gls{HBF} Gen-1 \gls{SLC} design point, and whether they serve the read or write path.
Five of the nine components are write-only and provide \emph{no} benefit for read-only weight access. 
The remaining read-side components either
become much smaller under burst-granular translation or move off the foreground path. 
This motivates an \gls{HBF} \gls{FTL} specialized for read-only, burst-granular access
rather than a direct import of \gls{SSD}-class write-path machinery.}

\begin{table}[h]
  \caption{\gf{Major \gls{SSD}-class \gls{FTL}
    components\revdelhpca{ and their relevance to an \gls{HBF} weight
    tier}.\revdelhpca{Garbage-collector bandwidth cost
    from~\cite{tavakkol2018mqsim}.}}}
  \centering
  \footnotesize
  \renewcommand{\arraystretch}{0.9}
  \setlength{\tabcolsep}{2.3pt}
  \begin{tabular}{@{} l cccc @{\hspace{7pt}} ccccc @{}}
  \toprule
  & \rotatebox{90}{\scriptsize\shortstack[l]{Page-level L2P\\(\SI{1}{\giga\byte} DRAM)}}
  & \rotatebox{90}{\scriptsize\shortstack[l]{ECC engine\\($<\!\SI{1}{\milli\meter\squared}$)}}
  & \rotatebox{90}{\scriptsize\shortstack[l]{Refresh sched.\\(counters)}}
  & \rotatebox{90}{\scriptsize\shortstack[l]{Bad-block map\\(\unit{KB} table)}}
  & \rotatebox{90}{\scriptsize\shortstack[l]{Garbage coll.\\(\SIrange{5}{30}{\percent} BW)}}
  & \rotatebox{90}{\scriptsize\shortstack[l]{Wear leveling\\(P/E ctrs.)}}
  & \rotatebox{90}{\scriptsize\shortstack[l]{Out-of-pl.\ buf.\\(\unit{MB} SRAM)}}
  & \rotatebox{90}{\scriptsize\shortstack[l]{Erase-blk.\ alloc.\\(\unit{KB} table)}}
  & \rotatebox{90}{\scriptsize\shortstack[l]{Write journal\\(\unit{MB} SRAM)}} \\
  \midrule
  \textbf{Read path}
  & \checkmark & \checkmark & \checkmark & \checkmark
  & & & & & \\
  \textbf{Write path}
  & \checkmark & \checkmark & \checkmark & \checkmark
  & \checkmark & \checkmark & \checkmark & \checkmark & \checkmark \\
  \bottomrule
  \end{tabular}
  \label{tab:ftl-components}
\end{table}

\section{\gf{\prop Overview}}
\label{sec:cmd_interface}


\gfhpca{\gfhpca{\prop is a workload-driven \gls{HBF} substrate whose goal is to sustain \gls{HBM}-side memory throughput for \gls{LLM} inference at low hardware and software overhead. 
The \emph{key idea} of \prop is to drive the \gls{HBF} access path from the \emph{dynamic} read stream the accelerator actually produces, rather than from a \emph{static} compile-time prediction, exploiting the burst-granular, read-only nature of weight access. 
\prop realizes this idea with three
mechanisms: 
\li~a hardware \emph{burst-buffer controller} (\cref{sec:bbc});
\lii~\emph{phantom-plane refresh} (\cref{sec:overview:refresh}); and
\liii~\emph{read-only \gls{FTL}} (\cref{sec:overview:ftl}).}}

\subsection{\gf{Burst-Buffer Controller}}
\label{sec:bbc}

\begin{figure}[ht]
    \centering
    \includegraphics[width=0.9\linewidth]{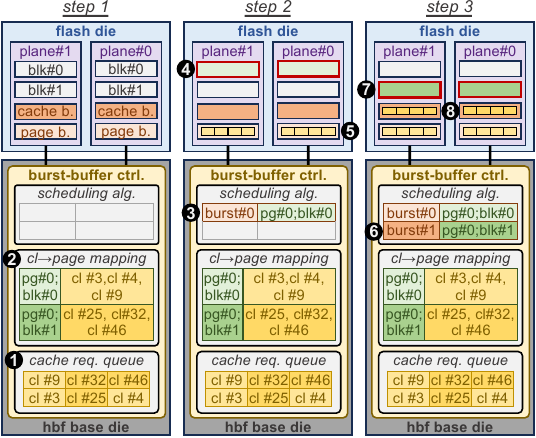}
    \caption{\gf{Overview of \prop burst-buffer controller.}}
    \label{fig:burst-buffer}
\end{figure}

\gf{The burst-buffer controller is a hardware unit in the \gls{HBF} base die that converts the accelerator's fine-grained cache line read stream into coarse-grained, plane-parallel \gls{HBF} bursts. 
Its goal is to keep the \gls{HBF} datapath continuously occupied without requiring an \gls{HBM}-side SRAM staging buffer or prefetch hints.
Fig.~\ref{fig:burst-buffer} shows the three main components of the
burst-buffer controller.
First, the \emph{cache request queue}~(\circled{1} in Fig.~\ref{fig:burst-buffer}) buffers cache line read requests to \gls{HBF}-resident weights after the \gls{HBM} base die's address decoder redirects them to the \gls{HBF} base die. 
Second, the \emph{cl-to-page mapping table}~(\circled{2})
groups queued cache line requests by their target \texttt{(page,
block)} pair, using the burst-granular translation maintained by the read-only \gls{FTL} (\cref{sec:overview:ftl}). 
The \emph{cl-to-page mapping table} bridges the granularity mismatch between accelerator cache line requests and the NAND flash 
page-read primitive, since a single page-read operation can serve multiple cache lines that reside in the same physical page. 
Third, the \emph{burst scheduler}~(\circled{3}) selects the next
\texttt{(page, block)} coordinate from the \emph{cl-to-page mapping table} using a predefined access-scheduling policy, e.g.,
first-ready, first-come-first-served~\cite{rixner2000memory,zuravleff1997controller}, and issues it as a \emph{burst}, i.e., a plane-parallel page read at the same \texttt{(page, block)} coordinate across every plane of every flash die.
\revdelhpca{This plane-parallel access amortizes one NAND page-read latency over a contiguous plane-fanned-out weight block, enabling the \gls{HBF} channel to sustain TB/s-class read bandwidth.}}

\gf{The burst-buffer controller operates in three main steps, as
shown in Fig.~\ref{fig:burst-buffer}. 
In the first step, cache line requests (from the accelerator) enter the \emph{cache request queue}~(\circled{1}) and are grouped in the \emph{cl-to-page mapping table}~(\circled{2}) according to their target \texttt{(page, block)} pair. 
In the example, six pending cache line requests collapse into two page-level groups, \texttt{(pg\#0,\,blk\#0)} and
\texttt{(pg\#0,\,blk\#1)}, each covering three requests. 
In the second step, the burst scheduler dequeues the first page-level group (following the pre-defined scheduling algorithm) and issues it as \texttt{burst\#0}, targeting \texttt{(pg\#0,\,blk\#0)}~(\circled{3}). 
Each plane activates \texttt{blk\#0} and senses \texttt{pg\#0} in
parallel~(\circled{4}), paying one NAND flash page-read latency to fill its page buffer with the plane-resident slice of
\texttt{burst\#0}~(\circled{5}). 
In the third step, while \texttt{burst\#0} drains from the page buffers to the \gls{HBM} base die over the \gls{D2D} link and then to the accelerator, the scheduler issues \texttt{burst\#1}, targeting
\texttt{(pg\#0,\,blk\#1)}~(\circled{6}). 
Each plane senses \texttt{blk\#1}'s \texttt{pg\#0} in parallel~(\circled{7}); the new data fills the page buffer, while \texttt{burst\#0}'s data moves into the cache buffer and continues draining~(\circled{8}). 
Because the page and cache buffers are distinct per-plane latches, the NAND flash sense for the next burst overlaps with the I/O drain of the current burst.}
\gf{In steady state, the controller presents one complete burst to the \gls{HBM} base die every drain interval, rather than every page-read latency plus drain interval. 
This is the timing condition required to reach peak \gls{HBF} read bandwidth. 
The pipeline is self-sustaining: the cl-to-page mapping table observes the pending request stream, and the scheduler can issue the next burst as soon as the current burst moves from the page buffers to the cache buffers. 
Thus, \prop exploits the existing per-plane page and cache buffers in \gls{HBF} to hide NAND page-read latency without compiler-inserted prefetches or a dedicated SRAM latency-hiding buffer.}


\subsection{\gf{Phantom-Plane Refresh}}
\label{sec:overview:refresh}

\gf{Phantom-plane refresh is a background-maintenance mechanism that
removes NAND flash refresh from the accelerator-visible read path.
The mechanism provisions one additional physical flash plane per flash
die and uses it as a rotating \emph{phantom-plane}.
\gfhpca{At any time, $N$ logical planes serve foreground flash reads
(from the accelerator), while the extra phantom-plane absorbs refresh
traffic in the background. Each block carries a read counter that trips
at the read-disturb threshold, with trip-points staggered so that
blocks fall due as a steady trickle rather than a storm
(Sec.~\ref{sec:motivation}); a due block of the current source plane
receives an \gls{ECC}-corrected copy, page by page, into a free block
of the phantom-plane. Once a block's copy completes, the
\emph{phantom-plane refresh controller} remaps that block in a
block-granular relocation table, erases the vacated block, and returns
it to the free-block pool. The vacated blocks of the source plane
accumulate into the next phantom-plane, so the phantom role rotates
round-robin through the physical planes and every plane is reprogrammed
once every $N{+}1$ rotation periods $T_{rot}$.}}
\revdelhpca{\gf{In comparison, in a baseline system, user reads and block refresh
share the same \gls{HBF} channel.
Refreshing one \gls{SLC} block costs one block erase plus one program
per page; with 1024 pages per block,
$t_\text{PROG}\!\approx\!\SI{50}{\micro\second}$ and
$t_\text{ERASE}\!\approx\!\SI{3}{\milli\second}$, a block refresh
takes tens of milliseconds, dominated by page programming.
If issued while user reads are pending, the refresh operation stalls
all reads to the refreshed block.
This stall is especially harmful for \gls{HBF}, whose read bandwidth
lies on the decode critical path (\cref{sec:motivation}).
Instead, phantom-plane refresh places copy, program, and erase work on
a parallel background-maintenance path, which allows the $N$ logical
planes to remain available for foreground reads, while the
phantom-plane absorbs the refresh traffic.}}

\begin{figure}[ht]
    \centering
    \includegraphics[width=0.75\linewidth]{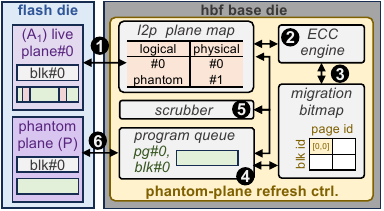}
    \caption{\gf{Phantom-plane refresh controller.}}
    \label{fig:phantom-refresh}
\end{figure}

\gf{Fig.~\ref{fig:phantom-refresh} shows the four new hardware
structures that compose the phantom-plane refresh controller (plus the
standard \gls{ECC} engine).
\gfhpca{First, the \emph{per-block refresh state} holds one read
counter per block, which triggers refresh at the staggered disturb
threshold, and the \emph{block-granular relocation table} with its
free-block pool, which records the physical block backing each logical
block. The controller updates the relocation table one block at a time
as copies complete, so every foreground read observes a consistent,
single-cycle mapping.}
\gfhpca{A small \emph{logical-to-physical (l2p) plane map} records which
physical plane currently holds the phantom role and masks it from the
read stripe.}
Second, the \emph{migration bitmap} tracks which \texttt{(block, page)}
entries of the source plane have already been copied into the
phantom-plane per-rotation.
The migration bitmap is required because \gls{LLM} inference rereads
hot weight pages many times within a rotation period, while each page
must be programmed into the phantom-plane only once.
Without avoiding duplication, the refresh path would try to program a
page at every read, causing either an overflow over the refresh path or
back-pressure foreground reads due to the mismatch between page read
and programming throughput.\footnote{Per plane, an \gls{SLC} read path
can deliver roughly
$\SI{4}{KB}/\SI{2}{\micro\second}\!\approx\!\SI{2}{GB/s}$, whereas the
program path absorbs only
$\SI{4}{KB}/\SI{50}{\micro\second}\!\approx\!\SI{80}{MB/s}$.}
Third, a \emph{program queue} decouples bitmap admission from NAND
flash programming. It accepts \gls{ECC}-corrected first-occurrence
pages from foreground reads and scrubber reads, and drains the
deduplicated stream into the phantom-plane at the slower per-plane
program rate.
Fourth, the \emph{scrubber} provides cold-page coverage. Foreground
reads opportunistically migrate hot pages into the phantom-plane, but
they may \emph{not} touch every \texttt{(block, page)} entry of the
source plane within a rotation period.
Therefore, the scrubber scans the migration bitmap for entries whose
bits remain clear and issues low-priority reads to those pages.
The returned pages are \gls{ECC}-corrected, inserted into the program
queue, and programmed into the phantom-plane.}

\gf{A phantom-plane rotation period proceeds in four main steps.
First, the phantom-plane refresh controller selects one live source
plane, e.g., $A_1$, and one phantom plane, $P$ (\circled{1} in
Fig.~\ref{fig:phantom-refresh}).
When a foreground request reads a page from $A_1$, the sensed data
reaches the \gls{HBF} base die and passes through the \gls{ECC} engine
(\circled{2}).
The corrected codeword then forks, and one copy returns to the
accelerator as the normal read response, while the other enters the
refresh path.
Second, the migration bitmap is consulted using the page's
\texttt{(block, page)} index (\circled{3}).
If the bit is clear, the controller sets it and enqueues the corrected
page for programming into the program queue (\circled{4}); if the bit
is already set, the refresh copy is dropped because that page has
already been migrated during the current rotation.
Third, before the completion of the rotation period, the background
scrubber walks the migration bitmap and issues low-priority reads for
entries whose bits remain clear (\circled{5})\gfhpca{; the scrubber
thereby also guarantees that the pass completes, and the data stays
error-free, even when the system is idle, and no \gls{HBF} reads
arrive}.
Fourth, when all bitmap entries are set, $P$ contains a complete
\gls{ECC}-correct copy of $A_1$ (\circled{6}).
\gfhpca{As each block's bitmap entries fill, the controller remaps that
block in the block-granular relocation table to its fresh copy in $P$,
erases the vacated block, and returns it to the free-block pool; when
all of $A_1$'s blocks have migrated, the drained plane's blocks form
the pool for the next rotation and $A_1$ becomes the next
phantom-plane (not shown).}}

\gf{\gfhpca{The per-plane program throughput sets the rotation step,
$T_\text{rot}\ge C_\text{plane}/\text{BW}_\text{prog}
=\SI{12.5}{\second}$, so a full pass over all $N{+}1$ planes completes
in about seven minutes, four orders of magnitude inside the \gls{SLC}
retention margin that cold blocks require. Hot blocks instead refresh
on their counters' schedule, and their worst measured demand of
\SI{0.86}{\giga\byte\per\second} per stack (Sec.~\ref{sec:motivation})
fits inside the \SI{1.28}{\giga\byte\per\second} that the sixteen
phantom-planes program.}}


\subsection{\gf{Read-Only \gls{FTL} Design}}
\label{sec:overview:ftl}

\gf{\prop uses a read-only \gls{FTL} to translate each logical \gls{LLM} weight-burst request into the physical \gls{HBF} \texttt{(block, page)} coordinate read by the burst-buffer controller. 
\revdelhpca{The design is specialized for the access pattern exposed by inference: weights are laid out once during model load, remain immutable on the accelerator read request path, and are later consumed as large, predictable bursts. 
This removes the need for the write-oriented mechanisms of a conventional \gls{SSD} \gls{FTL}. 
Therefore, the foreground path reduces to a deterministic SRAM lookup, rather than a page-level translation followed by garbage collection, wear leveling, and metadata updates.}}
\revdelhpca{\gf{A conventional \gls{SSD}-class \gls{FTL} is designed for arbitrary host writes. 
It maintains page-granular logical-to-physical mappings, performs out-of-place updates, reclaims invalidated space through
garbage collection, and spreads P/E cycles through wear leveling. 
On an \gls{HBF} stack, this design would be both too large and too disruptive. 
For example, a \SI{512}{GB} stack with \SI{4}{KB} pages contains roughly 128\,M pages; at four to eight bytes per mapping
entry, the translation table alone would consume \SIrange{512}{1024}{MB} of metadata. 
More importantly, garbage collection and wear leveling would share the same \gls{HBF} channel as foreground weight reads, introducing background traffic and unpredictable stalls on the decode critical path.}}
\gf{Concretely, \prop employs a single \emph{burst translation table} in the \gls{HBF} base die. 
Each entry maps a logical burst index to a physical \texttt{(block, page)} coordinate. 
With \SI{2}{MB} bursts on a \SI{512}{GB} stack, the table contains 256\,K entries and occupies about \SI{1}{MB} of base-die SRAM, several orders of magnitude smaller than a page-granular \gls{FTL}. 
The table is populated once from the compiler-emitted weight layout during model load and remains constant throughout inference. 
Thus, every foreground access performs a single-cycle SRAM lookup and obtains the physical coordinate used by the burst-buffer controller to issue the corresponding plane-parallel
read.}
\gf{The responsibilities normally handled by a full \gls{FTL} are
either unnecessary or moved off the foreground path. 
Out-of-place updates and garbage collection can be avoided because weights are \emph{not} modified during inference.
Wear leveling is provided implicitly by phantom-plane refresh (\cref{sec:overview:refresh}), which rotates the phantom role and reprograms planes uniformly over time. 
The temporary exclusion of the current phantom plane is handled by the logical-to-physical plane map maintained by the refresh controller. 
\revdelhpca{As a result, \prop retains only the translation state needed for read-only burst access, eliminating large metadata structures, metadata writes, and background \gls{FTL}-induced contention during inference.}}

\subsection{\gf{Write Path: Model Load and Persistence}}
\label{sec:mech-write}

\revdel{\gf{\prop keeps NAND flash writes off the foreground inference path. 
The only bulk write occurs during \emph{model load}, when the runtime programs \gls{HBF} with model weights and initializes the burst translation table (\cref{sec:overview:ftl}). 
After deployment, \gls{HBF}-resident weights are immutable, so no foreground inference request triggers a NAND program or erase. 
After a power cycle, \prop performs only \emph{metadata recovery}: it reloads reconstructable base-die state from a reserved, \gls{ECC}-protected metadata region in \gls{HBF}.}}

\paratitle{\gf{Model Load and Address Striping}}
\gf{The runtime writes the model once at deployment time using a fixed mapping function that places each \SI{2}{\mega\byte} burst in the physical layout expected by the burst-buffer controller. 
The layout
ensures that one burst can be fetched by a single plane parallel NAND sense.}
\gf{An \gls{HBF} Gen-1 \gls{SLC} stack with capacity $C=\SI{512}{\giga\byte}$ and burst size $B=\SI{2}{\mega\byte}$ contains $N_\text{burst}=C/B=2^{18}$ bursts. 
Each burst is striped across all $D{\times}P=16{\times}32=512$ planes $P$ of the \gls{HBF} stack with $D$ dies, one \SI{4}{\kilo\byte} slice per plane. 
For burst index $b\in[0,N_\text{burst})$ and
$N_\text{pg/blk}=1024$ pages per block, the common physical coordinate is $\operatorname{block}(b)=\lfloor b/N_\text{pg/blk}\rfloor$ and $\operatorname{page}(b)=b\bmod N_\text{pg/blk}$. Slice $s\in[0,512)$ of the burst is placed at this same \texttt{(block, page)} coordinate in $\operatorname{die}(s)=\lfloor s/P\rfloor$ and $\operatorname{plane}(s)=s\bmod P$.
Thus, each logical burst maps to one physical \texttt{(block, page)} coordinate that is shared across all 512 planes, while the slice-to-plane assignment is implicit. 
The burst translation table (\cref{sec:overview:ftl}) stores only the compact \texttt{(block, page)} coordinate for each logical burst index.}

\revdelhpca{\gf{The runtime programs the burst slices using the per-plane program primitive, whose throughput is
$\SI{4}{\kilo\byte}/t_\text{PROG}=\SI{80}{\mega\byte\per\second}$ for
$t_\text{PROG}=\SI{50}{\micro\second}$. 
Conservatively assuming one active program stream per die, the 16 dies in a stack provide an aggregate write bandwidth of
$16{\times}\SI{80}{\mega\byte\per\second}
=\SI{1.28}{\giga\byte\per\second}$.
Loading a \SI{512}{\giga\byte} stack thus takes
$\sim\!\SI{6.7}{\minute}$; an eight-stack \SI{4}{\tera\byte} deployment takes the same time when all stacks load in parallel. 
This deployment-time cost is paid once and amortized over
the lifetime of the loaded model.}}

\paratitle{\gf{Power-Loss Recovery}}
\gf{Although \gls{HBF} is non-volatile, \prop does \emph{not} rely on preserving resident weights across an unplanned power loss. 
The outage duration is unbounded, and guaranteeing correctness after arbitrary retention drift would require persistent refresh state, read-disturb counters, and recovery metadata. 
Maintaining this state would reintroduce write-path support that \prop deliberately keeps off the foreground inference path. 
Instead, \prop treats every power-up as a clean installation (which we estimate as taking $t_\text{erase}+t_\text{reload}\approx\SI{6.7}{\minute}$).}

\section{\gf{Methodology}}
\label{sec:methodology}



\gfhpca{We evaluate \prop with an in-house trace-driven simulator that models the five tiers a weight read crosses, namely the \glspl{SM}, the on-die L2 cache, \gls{HBM}, the staging buffers, and \gls{HBF}. 
It consumes a per-\gls{SM} stream of compute, memory, communication, and synchronization events produced by the WSE workload model~\cite{wseworkload}. In multi-package configurations, it charges \emph{every} transfer that crosses a GPU package boundary and stalls each GPU package at a layer barrier until its peers arrive, so the multi-package baselines pay their interconnect and synchronization costs.}
\gfhpca{We calibrate our \gls{HBM} model against Ramulator~2.0~\cite{luo2023ramulator}, driving it with the address streams from our \gls{LLM} inference traces spanning four batch sizes and four context lengths. 
We observe from Ramulator that the sustained-to-peak \gls{HBM} bandwidth is
\SIrange{67.8}{69.9}{\percent} and does \emph{not} vary with model, batch size, or context length.}
\gfhpca{We faithfully model a B200-class accelerator~\cite{nvidia2025b200}, deriving arithmetic from the tensor shapes the trace carries and executing it at the product of peak throughput and a machine-utilization factor. Resolving arithmetic analytically and
spending fidelity on the memory system is the same approach that recent
academic \gls{LLM}-inference simulators take~\cite{park2024attacc}.
Our energy model charges memory and inter-GPU traffic per bit while \gls{GPU} arithmetic is computed per operated byte.
Table~\ref{tab:hw-params} lists our system parameters.
}

\begin{table}[!h]
  \caption{Simulated system configuration.}
  \label{tab:hw-params}
  \centering
  \footnotesize
  \renewcommand{\arraystretch}{1.1}
  \begin{tabularx}{\columnwidth}{@{}l>{\raggedright\arraybackslash}X@{}}
    \toprule
    \textbf{GPU}          & B200 (Blackwell)~\cite{nvidia2025b200}; \num{148}
                            \glspl{SM} at \SI{2.5}{\giga\hertz}; \num{11000} fp16
                            MACs/\gls{SM}/cycle; MFU \num{0.5}; L2 \SI{126}{\mega\byte}
                            at \SI{21}{\tera\byte\per\second} \\
    \midrule
    \textbf{HBM3e}        & \SI{192}{\giga\byte},
                            $8\times\SI{1}{\tera\byte\per\second}$
                            stacks~\cite{jedec2022hbm3}; calibrated \num{0.70}
                            sustained/peak; row hit/miss \num{70}/\SI{105}{\nano\second};
                            \SI{20}{\giga\byte} KV reserve \\
    \midrule
    \textbf{HBF}          & Gen-1 \gls{SLC}~\cite{sandisk2025hbf,ha2026h3};
                            \SI{4}{\tera\byte}, $8\times$(\SI{512}{\giga\byte},
                            \SI{1}{\tera\byte\per\second}) stacks; \num{16} dies/stack,
                            \num{32} planes/die, \num{256} blocks/plane,
                            $\num{1024}\times\SI{4}{\kilo\byte}$ pages/block;
                            $t_R/t_\text{PROG}/t_\text{ERASE}=
                            \SI{2}{\micro\second}/\SI{50}{\micro\second}/\SI{3}{\milli\second}$ \\
    \midrule
    \textbf{Reliability}  & read-disturb \num{1e6} senses/block; \num{1e5} P/E
                            cycles/block; $>\!1$~yr end-of-life retention; in-place
                            block refresh \SI{54.2}{\milli\second} over \num{512}
                            planes/stack \\
    \midrule
    \textbf{Staging}      & \gls{LHB} \SI{2}{\mega\byte}/slot, \num{5000}-cycle
                            min.\ residency; \SI{2}{\mega\byte} burst (\num{512}
                            page-senses); weights block-striped across stacks;
                            \gls{D2D} \SI{1}{\tera\byte\per\second}/stack~\cite{sharma2024universal} \\
    \midrule
    \textbf{Interconnect} & NVLink, $18\times\SI{450}{\giga\byte\per\second}$ per
                            direction per \gls{GPU}; \SI{800}{\nano\second}/hop;
                            per-layer barrier convergence \\
    \midrule
    \textbf{SSD}          & PCIe Gen5\,$\times$\,4~\cite{tavakkol2018mqsim};
                            \SI{14}{\giga\byte\per\second}; \SI{40}{\micro\second} read \\
    \midrule
    \textbf{Energy}       & \gls{GPU} \SI{1000}{\watt} TDP / \SI{400}{\watt} sync
                            spin / \SI{100}{\watt} idle; ALU
                            \SI{0.32}{\pico\joule\per\byte}~\cite{park2024attacc};
                            \gls{LHB} leakage \SI{50}{\milli\watt}; per bit:
                            \gls{HBM} \num{4}, \gls{HBF} read \num{20} (swept
                            \numrange{10}{25}), \gls{HBF} program \num{100},
                            \gls{D2D} \num{1}, NVLink \num{1.3}~\cite{park2024attacc},
                            \gls{SSD} \num{100}~\si{\pico\joule\per\bit} \\
    \bottomrule
  \end{tabularx}
\end{table}

\paratitle{Workloads}
\gfhpca{We evaluate \prop on six production \glspl{LLM}, including five \gls{MoE} models, namely DeepSeek-V3~\cite{liu2024deepseek}, DeepSeek-V4-Pro~\cite{deepseekai2026deepseekv4}, Qwen3-235B-A22B~\cite{bai2025qwen3}, Llama-4 Maverick~\cite{meta2025llama}, and Kimi~K2~\cite{team2025kimi}, and a dense model, Llama~3.1-405B~\cite{dubey2024llama3}.
Each model runs at its native data precision.
We generate full-model decode traces with the WSE workload generator~\cite{wseworkload} at four batch sizes, $bsz\!\in\!\{1,4,16,64\}$, and four context lengths, $\{2,8,32,128\}\,\text{K}$~tokens.
We report results at the largest evaluated context length, otherwise specified.}
\gfhpca{We capture routed-expert distributions for our five \gls{MoE} models from a \emph{real} \gls{GPU} deployment serving MMLU~\cite{hendrycks2020measuring} prompts, and drive \emph{every} \gls{MoE} trace with the measured distributions.}

\paratitle{Baselines}
\gfhpca{We compare \prop against three deployments.
\li~\emph{HBM+SSD} keeps one \gls{GPU} and spills weights to an NVMe \gls{SSD} once they exceed the package's \SI{192}{\giga\byte} of \gls{HBM}.
\lii~\emph{HBM-only} holds the whole model in aggregate \gls{HBM} with \emph{no} \gls{HBF} tier, sharded across the fewest packages that hold the weights alone; where the KV cache exceeds the memory capacity a package has left, we annotate the cell as \emph{capacity-infeasible} rather than re-sharding it.
\liii~\emph{H$^3$}~\cite{ha2026h3} keeps one \gls{GPU} package and stages bursts in two \gls{LHB} slots with FIFO eviction, issued from compiler-emitted layer-ahead hints. 
We emit a prefetch hint \emph{only} where a compiler could know the address before the gating layer runs.}

\textbf{\section{\gf{Evaluation}}
\label{sec:eval}}

\subsection{\gls{LLM} Inference Analysis}

\begin{figure}[ht]
  \centering
  \includegraphics[width=\linewidth]{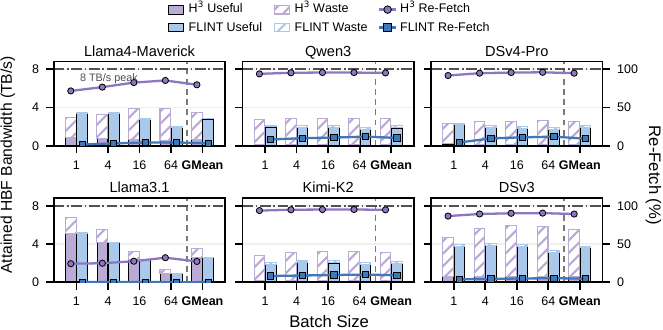}
  \caption{\gfhpca{Attained \gls{HBF} bandwidth (left y-axis, bars) and re-fetch percentage (right y-axis, lines) for H$^{3}$ and \prop.}}
  \label{fig:stall-bw}
\end{figure}

\paratitle{\gls{HBF} Bandwidth}
\gfhpca{Fig.~\ref{fig:stall-bw} shows the attained \gls{HBF} bandwidth, decomposed into useful and re-fetched traffic (left y-axis, bars), and the re-fetch share of all \gls{HBF} traffic (right y-axis, lines) for H$^{3}$ and \prop across six models and batch sizes $\{1,4,16,64\}$ at 128K
context. We make three observations.}
\gfhpca{First, \prop converts the \gls{HBF} channel into useful weight bandwidth, because it holds each fetched burst in the \gls{HBF}'s page and cache buffers until the
\glspl{GPU} drain it. 
\prop consumes \SIrange{90}{97}{\percent} of the \gls{HBF} traffic it fetches and sustains \SIrange{1.9}{3.6}{\tera\byte\per\second} of useful bandwidth on the \gls{MoE} models, $4.0\times$--$14.3\times$ the useful bandwidth of H$^{3}$ ($6.2\times$ on average across all six models).
Second, H$^{3}$ moves $1.2\times$--$1.5\times$ \emph{more} raw \gls{HBF}
bytes than \prop yet re-fetches \SIrange{80}{95}{\percent} of them on the \gls{MoE} models, because expert-weight bursts leave its \gls{LHB} before every consuming \gls{GPU} reads them, so the controller reads the same
NAND pages again.
Third, dense Llama~3.1-405B is the boundary case: each decode token reads
every weight exactly once in layer order, so H$^{3}$'s re-fetch share falls to \SI{27}{\percent} and both controllers deliver the same \SI{2.6}{\tera\byte\per\second} of useful bandwidth\revdelhpca{, with H$^{3}$ recovering its loss by moving $1.4\times$ \prop's raw traffic}. 
\revdelhpca{We conclude that \prop turns the \gls{HBF} channel into useful weight bandwidth, whereas H$^{3}$ spends most of the same channel re-reading bursts it already fetched.}}

\begin{figure}[ht]
  \centering
  \includegraphics[width=\linewidth]{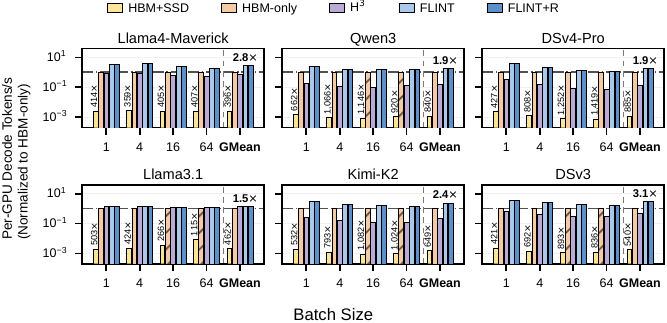}
  \caption{\gfhpca{Per-GPU decode throughput.}}
  \label{fig:throughput}
\end{figure}

\paratitle{Performance Analysis}
\gfhpca{Fig.~\ref{fig:throughput} shows per-\gls{GPU} decode throughput for five configurations across six models and batch sizes $\{1,4,16,64\}$ at 128K context, normalized to the \gls{HBM}-only baseline. 
Hatched bars mark batches whose KV cache \emph{no} longer fits the min-fit \gls{HBM}-only configuration\revdelhpca{ (their values are excluded from the reported geometric means)}. 
\revdelhpca{For fairness, we do \emph{not} charge H$^{3}$ with refresh cost.}
We make four observations.}
\gfhpca{First, \prop achieves $1.5\times$--$3.1\times$ the per-\gls{GPU} decode throughput of \gls{HBM}-only ($2.2\times$ across all six models; up to $3.7\times$ at $bsz{=}1$), because in the \gls{HBM}-only baseline, the slowest \gls{GPU} package dictates execution time: we observe that GPUs spend \SIrange{74}{85}{\percent} of decode time waiting at layer barriers at $bsz{=}1$.
In contrast, \prop serves the same model from a single package with \emph{no} cross-\gls{GPU} synchronization. 
At larger batch sizes, the \gls{HBM}-only baseline can amortize such synchronization overheads (synchronization drops to \SI{22}{\percent} at $bsz{=}64$), which narrows the performance gap with \prop.
However, at 128K context length, the aggregate capacity of the min-fit deployment \emph{fails} to hold both weights and KV cache for several models (i.e., Qwen3, Llama3.1, Kimi-K2, and DSv3), which would force the adoption of more \gls{GPU} packages.
Second, \prop achieves $4.0\times$--$14.3\times$ the per-\gls{GPU}
decode throughput of H$^{3}$ on the \gls{MoE} models ($6.2\times$ on average across all six models), matching the useful-\gls{HBF}-bandwidth ratios of Fig.~\ref{fig:stall-bw}.
\revdelhpca{This performance improvement happens because decode time is set by how fast the active weights stream out of \gls{HBF}: H$^{3}$ spends \SIrange{80}{95}{\percent} of its channel re-fetching evicted bursts.}
Third, read-disturb refresh does \emph{not} change \prop's performance: \prop with phantom-plane refresh (\prop+R) delivers throughput \emph{identical} to \prop in every $(\textit{model}, bsz)$ cell\revdelhpca{, because refresh rebuilds each due block in a reserved spare plane off the live read stripe, so the read path never observes it}. 
Fourth, \gls{HBM}+\gls{SSD} is two to three orders of magnitude slower per \gls{GPU} than every on-package-memory configuration ($396\times$--$885\times$ below the \gls{HBM}-only baseline on average)\revdelhpca{, because the \SI{14}{\giga\byte\per\second} \gls{SSD} link stretches every decode step; \gls{SSD}-backed inference is \emph{not} a practical operating point at frontier-model scale}.
\revdelhpca{We conclude that \prop is the only evaluated single-\gls{GPU}
configuration that exceeds the per-\gls{GPU} decode throughput of the capacity-sharded \gls{HBM}-only cluster at frontier-model scale, and that it does so with read-disturb refresh active at no throughput cost.}}

\begin{figure}[ht]
  \centering
  \includegraphics[width=\linewidth]{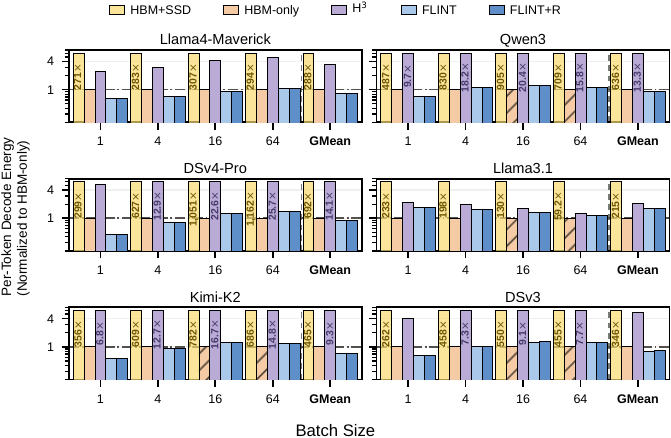}
  \caption{\gf{Per-token decode energy.}}
  \label{fig:energy}
\end{figure}

\paratitle{Energy Analysis}
\gfhpca{Fig.~\ref{fig:energy} shows the per-token decode energy of five configurations, normalized to the \gls{HBM}-only baseline\revdelhpca{, across six models and batch sizes $\{1,4,16,64\}$ at 128K context}.
We make five observations.}
\gfhpca{First, \prop consumes $0.72\times$--$0.90\times$ the energy per-token of the \gls{HBM}-only baseline, on average across the five \gls{MoE} models.
This happens because the HBM-only baseline spends \SIrange{17}{64}{\percent} of its energy busy-waiting at layer barriers of multiple \gls{GPU} packages, while \prop avoids that by operating within a single package.
The energy saving is largest at $bsz{=}1$, where \prop consumes \emph{only} $0.45\times$--$0.73\times$ the HBM-only energy per-token (a $2.2\times$ reduction on DeepSeek-V4-Pro), and decreases at larger batch sizes as the HBM-only baseline amortizes its barrier waits over more tokens.
Second, we observe an energy increase at the dense Llama model of $1.64\times$ that of HBM-only execution.
This happens because \prop re-streams every weight from \gls{HBF} at a higher per-bit read energy than the HBM-only baseline.
Third, refresh (i.e., FLINT+R) adds at most \SI{0.31}{\percent} of refresh program energy.
Fourth, against H$^{3}$, \prop reduces energy by $4.2\times$--$15.7\times$ on the \gls{MoE} models ($1.2\times$ on dense Llama), because H$^{3}$'s high re-fetched \gls{HBF} traffic (Fig.~\ref{fig:stall-bw}) increases
\li~overall \gls{HBF}-side energy consumption for extra NAND read operations, and
\lii~\gls{GPU} cycles due to longer waits for weight traffic from \gls{HBF}.
Fifth, \gls{HBM}+\gls{SSD} is far more energy-intensive than every on-package-memory configuration, consuming $215\times$--$692\times$ the \gls{HBM}-only energy (on average)\revdelhpca{, because the
\SI{14}{\giga\byte\per\second} \gls{SSD} link stretches decode wall
time by orders of magnitude and the package's power integrates over
that wall.
We conclude that \prop reduces decode energy by fitting the \gls{LLM} on a single package and converting the \gls{HBF} channel into useful weight bandwidth}.}

\begin{figure}[ht]
  \centering
  \includegraphics[width=\linewidth]{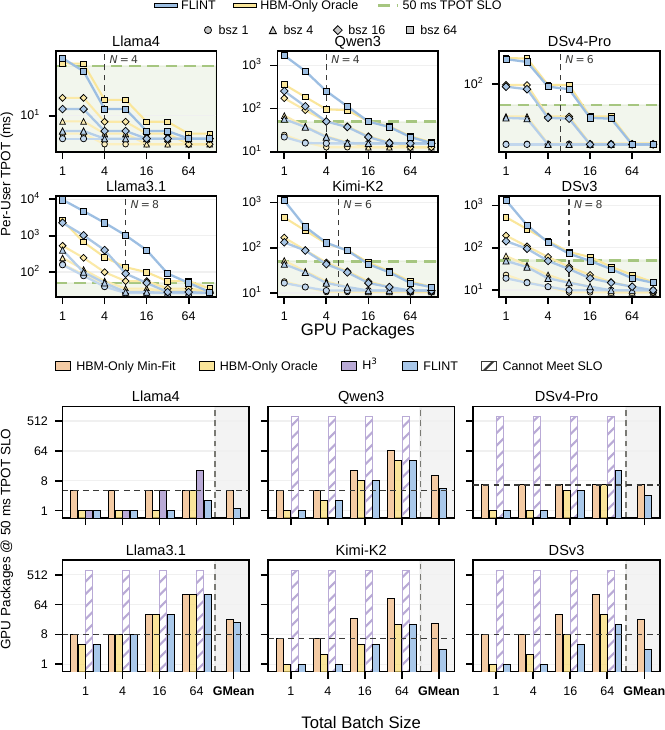}
  \caption{\gfhpca{Capacity scaling at 128K context size. (top)~Per-user TPOT vs.\ GPU packages, \prop (blue) against the capacity-lifted \gls{HBM} oracle (yellow). (bottom)~GPU packages needed to serve a total decode batch within the SLO; dashed lines mark the \gls{HBM} min-fit GPU-package count $N$.}}
  \label{fig:capacity-scaling}
\end{figure}

\paratitle{Capacity Scaling}
\gfhpca{Fig.~\ref{fig:capacity-scaling} evaluates how many GPU packages each system needs to serve \gls{LLM} decode at 128K context size. 
We evaluate four deployment options:
\li~\emph{HBM min-fit} shards the model across the $N$ GPU packages whose combined \gls{HBM} holds the weights, and replicates that shard until the replicas meet the target SLO; 
\lii~the \emph{capacity-lifted oracle} removes the \gls{HBM} capacity constraint and serves both weights and KV cache from \gls{HBM} at any GPU-package count; 
\liii~H$^{3}$ and
\liv~\prop place the weights on \gls{HBF} and shard across any measured GPU-package count. 
For every deployment option and total batch size, we scale the GPU-package count until the deployment meets the \SI{50}{\milli\second} \gls{TPOT} \gls{SLO}. 
\revdelhpca{A deployment shards the model across $k$ GPU packages and replicates that shard $r$ times; each replica serves an equal share of the batch.
We report the smallest total count, $r\times k$ GPU packages, that meets the \gls{SLO}.} 
We make three observations.}
\gfhpca{First, \prop follows the capacity-lifted oracle across the measured range and surpasses it on \emph{every} model at $bsz{=}1$ on a single GPU package, and on DSv3 from four GPU packages onward. 
\prop surpasses the capacity-lifted oracle because the oracle serves weight and reads from the same \gls{HBM} channel, while \prop streams weights over the otherwise-idle \gls{HBF} channel in parallel with the \gls{HBM} KV reads and serializes only at the D2D links. 
\revdelhpca{\prop falls behind the capacity-lifted oracle only where its KV cache spills to \gls{HBF}, at low GPU-package counts and large batch sizes, and the gap closes as sharding increases the per-package \gls{HBM} capacity left
for the KV cache.}
Second, at $bsz{=}\{1,4\}$, \prop serves five of the six models
with \emph{one} or \emph{two} GPU packages. 
At $bsz{=}1$, a single \prop GPU package
serves each of the five \gls{MoE} models within the SLO, while the \gls{HBM} min-fit needs four to eight GPU packages only to hold the weights. At $bsz{=}4$, one \prop GPU package still suffices for four of the five \gls{MoE} models, and two for Qwen3. Dense Llama needs four \prop GPU packages at $bsz{=}1$ and eight at $bsz{=}4$.
Third, at $bsz{=}64$, \prop needs fewer GPU packages than the \gls{HBM} min-fit on four of the six models (2 vs.\ 4 on Maverick, 32 vs.\ 64 on Qwen3, 16 vs.\ 96 on Kimi-K2, and 16 vs.\ 128 on DSv3), for the same reason it follows the capacity-lifted oracle. 
Dense Llama ties at 128 GPU packages, because its throughput, not its weight capacity, sets the GPU-package count. The single configuration where \prop needs more GPU packages than the \gls{HBM} min-fit is DSv4-Pro at $bsz{=}64$, where \prop's best shard misses the SLO by \SI{0.7} {\milli\second} (50.7 vs.\
\SI{50}{\milli\second}) and the hard threshold forces 16 replicated GPU packages against the min-fit's six. 
\revdelhpca{In deployment, this margin disappears because DSv4-Pro decodes with a multi-token-prediction head that drafts one extra token per step, and accepting more than $1.4\%$ of its drafts returns the six-package deployment to the SLO.
We conclude that \prop yields a substantially cheaper deployment, meeting the SLO with $3.1\times$ fewer GPU packages than the \gls{HBM} min-fit on average over the 24 measured (model, batch) configurations ($3.8\times$ over the five \gls{MoE} models, and up to $8\times$ on DSv3).}}

\subsection{\gf{\gls{HBF} Refresh \& Lifetime Analysis}}
\label{sec:eval-lifetime}

\begin{figure}[ht]
  \centering
  \includegraphics[width=\linewidth]{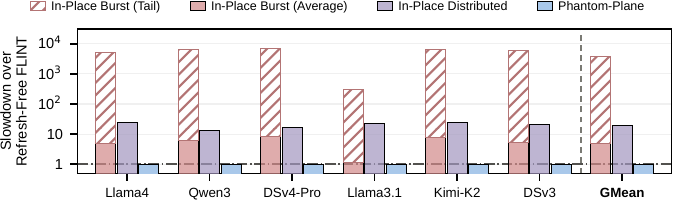}
  \caption{\gfhpca{Decode throughput overhead due to read-disturb-induced refresh operations.}}
  \label{fig:refresh-cost}
\end{figure}

\paratitle{Refresh Execution Cost}
\gfhpca{Fig.~\ref{fig:refresh-cost} shows the impact of \gls{HBF} refresh on \gls{LLM} decode throughput for three refresh mechanisms: 
\li~\emph{in-place burst refresh}, 
\lii~\emph{in-place distributed refresh}, and 
\liii~\prop's off-path phantom-plane refresh.
We use a 128K context size and $bsz{=}1$.
We make three observations.
First, \prop's off-path refresh execution incurs \emph{no} refresh overhead: \prop matches the decode throughput of a refresh-free operation, since the target refresh block is erase-and-program outside the foreground read stream.
Second, in-place refresh costs $20\times$ on average ($13\times$--$24\times$ across models), because a plane is a serial resource and every in-place refresh blocks one of the 32 serving planes for \SI{54}{\milli\second}, about $5{,}000$ read slots.
Third, in-place burst refresh concentrates its cost in one token, since the counters are initialized together, thousands of blocks come due at the same time, and decode stalls $3{,}700\times$ on average ($303\times$--$6{,}771\times$) at the token where their refreshes land, while the average over the whole refresh period is $4.8\times$ ($1.2\times$--$8.5\times$).
\revdelhpca{We conclude that \prop's phantom-place refresh is an effect mechanism to mitigate read-disturb-triggered refresh operations in \gls{LLM} inference.}}

\begin{figure}[ht]
  \centering
  \includegraphics[width=\linewidth]{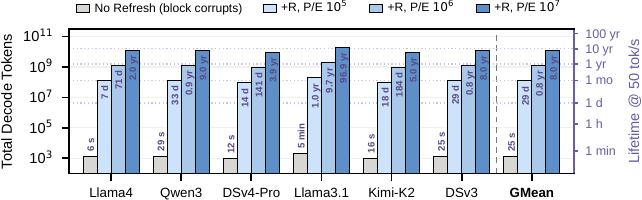}
  \caption{\gfhpca{Total decode tokens \prop serves before the always-on block wears out, per \gls{SLC} P/E budget (left y-axis), and the corresponding lifetime at a 50-tok/s decode rate (right y-axis).}}
  \label{fig:lifetime}
\end{figure}

\paratitle{Lifetime}
\gfhpca{Fig.~\ref{fig:lifetime} shows how long an always-on weight block (one that decode reads every token) survives under each model's 128K context size at $bsz{=}1$ (left y-axis) and the corresponding lifetime at a 50-tok/s decode rate.
Since vendors have \emph{not} yet published an \gls{HBF} endurance figure, we sweep the P/E budget
from the commodity-\gls{SLC} floor of $10^{5}$ cycles to a projected $10^{7}$.
We make two observations.
First, \prop's refresh mechanism \emph{significantly} increases device lifetime.
A block that corrupts within 25 seconds \emph{without} refresh survives an average of 
\li~29 days of continuous decode at $10^{5}$ P/E cycles (7--33 days on the \gls{MoE} models and 1.0 year on dense Llama), 
\lii~0.8 years at $10^{6}$ (71 days--9.7 years), and 
\liii~8.0 years at a projected $10^{7}$ (2.0--96.9 years).
Second, by reducing re-fetching, \prop's burst-buffer controller contributes to an increase in device lifetime.
Since popular experts concentrate flash reads, we observe that the most-read block refreshes $1.1\times$--$7.9\times$ more often than the average one and wears out sooner. \prop's coalescing bounds the skew read distribution, because efficiently merging duplicate page requests leads to the most popular block to be read at most about once per token.
\revdelhpca{We conclude that phantom-plane refresh shifts the \gls{HBF} weight tier's failure mode from read-disturb corruption within seconds-to-minutes to program wear over months-to-decades, scaling linearly with the \gls{SLC} P/E budget.}}

\subsection{\gf{Area Analysis}}
\label{sec:eval-area}

\gf{We estimate \prop's area overhead in two parts: additional storage
inside each \gls{HBF} die, and additional control/storage structures in
the \gls{HBF} base die.
We model all added SRAM structures with
CACTI~\cite{cacti,muralimanohar2007optimizing} at \SI{22}{\nano\meter},
the smallest valid SRAM node in CACTI~7, and project to
\SI{7}{\nano\meter} by dividing by the area-scaling factor reported
by~\cite{stillmaker2017scaling}.
We estimate synthesized control logic from comparable memory-controller
designs scaled to the same node. \gfhpca{We quote per-structure areas at \SI{22}{\nano\meter}, where CACTI models them, and report totals at the projected \SI{7}{\nano\meter} node.}}
\gfhpca{Table~\ref{tab:area} breaks down \prop's area cost. }

\begin{table}[!h]
  \caption{\gfhpca{\prop{} area overhead.}}
  \label{tab:area}
  \centering
  \footnotesize
  \setlength{\tabcolsep}{4pt}
  \renewcommand{\arraystretch}{0.9}
  \begin{tabularx}{\columnwidth}{@{}>{\raggedright\arraybackslash}X r@{}}
    \toprule
    \textbf{Structure (size)} & \textbf{Area} \\
    \midrule
    \emph{\gls{HBF} die:} phantom-plane, 1 plane per die & $+3.1\%$ \\
    \midrule
    \multicolumn{2}{@{}l}{\emph{\gls{HBF} base die}
      (\si{\milli\meter\squared} at \SI{22}{\nano\meter}):}\\
    \quad Burst translation table (\SI{1}{\mega\byte}) & 1.29 \\
    \quad Cache-request queue $+$ cl-to-page map (\SI{256}{\kilo\byte}) & 0.35 \\
    \quad Migration bitmaps ($16{\times}\SI{32}{\kilo\byte}$) & 0.71 \\
    \quad Program queues ($16{\times}\SI{0.5}{\mega\byte}$) & 10.5 \\
    \quad Read counters $+$ relocation table (\SI{0.83}{\mega\byte}) & 1.15 \\
    \quad Synthesized control logic & $<$0.3 \\
    \quad \textbf{Total} (\textbf{\SI{3.9}{\milli\meter\squared}}
      at \SI{7}{\nano\meter}) & \textbf{14.3} \\
    \midrule
    Page-level \gls{FTL} table (\SI{512}{\mega\byte}), for reference
      & 180 at \SI{7}{\nano\meter} \\
    \bottomrule
  \end{tabularx}
\end{table}

\section{Related Work}
\label{sec:related}


\gf{To our knowledge, \prop is the first \gls{HBF} substrate that sustains high \gls{HBF} bandwidth on the \gls{LLM} decode critical path without a dedicated \gls{HBM}-side SRAM staging buffer or compiler-emitted prefetch hints. \prop co-designs a runtime-aware burst-buffer controller, phantom-plane refresh, and a burst-granular read-only \gls{FTL} for immutable \gls{LLM} weight access. In this section, 
we contrast \prop with prior work along two themes.}

\paratitle{\gls{HBF}-Based Systems}
\gf{Prior \gls{HBF}-based systems for \gls{LLM} inference~\cite{ha2026h3,suh2026hbfworkload,kim2026hbfroadmap,hsu2026haven} commonly employ 
\li~an \gls{HBM}-side SRAM staging buffer that holds prefetched weight bursts, and 
\lii~compiler- or programmer-emitted hints fill this buffer at
layer boundaries. \revdelhpca{H$^{3}$~\cite{ha2026h3} introduces the dual-buffered \gls{LHB}
organization that subsequent \gls{HBF} works inherit;
KAIST~\cite{suh2026hbfworkload,kim2026hbfroadmap} refines the
workload model and \gls{HBF} roadmap parameters; and
HAVEN~\cite{hsu2026haven} applies the same staging-buffer template to vector-search retrieval. 
These designs keep the staging buffer plus static prefetch architecture intact and do \emph{not} address read-disturb
refresh on a loaded foreground channel.} 
Our analysis shows that this organization \emph{cannot} saturate the \gls{HBF} channel bandwidth, since static issue memory request order does \emph{not} reliably match the runtime memory access order under \gls{MoE} routing and attention reuse.} \gf{Other \gls{HBF}-related works~\cite{wu2026memexplorer, sun2025lincoln} are orthogonal to \prop.
MemExplorer~\cite{wu2026memexplorer} explores heterogeneous memory systems that include \gls{HBF} as one possible capacity tier but does \emph{not} propose new \gls{HBF} control mechanisms. 
\prop provides such a mechanism and raises the attainable bandwidth of the \gls{HBF} tier toward the \gls{HBF} bandwidth channel ceiling assumed by~\cite{wu2026memexplorer}.
Lincoln~\cite{sun2025lincoln} targets LPDDR-interfaced compute-enabled flash and near-flash compute. 
In contrast, \prop targets \gls{HBF} attached through the
\gls{HBM}-side \gls{D2D} link, requires \emph{no} in-flash compute, and obtains its bandwidth from plane-parallel burst staging through unmodified NAND flash dies.}

\paratitle{\gf{Non-Volatile Memory for \gls{LLM} Inference}} \gf{Several prior works~\cite{sheng2023flexgen,alizadeh2024llmflash,aminabadi2022deepspeed,xue2024powerinfer2,wang2024ripple,jia2025activeflow} integrate non-volatile memory into \gls{DNN} and \gls{LLM} inference systems at several layers of the stack. 
\revdelhpca{Offload systems, such as FlexGen~\cite{sheng2023flexgen},
LLM-in-a-Flash~\cite{alizadeh2024llmflash}, DeepSpeed-Inference~\cite{aminabadi2022deepspeed},
PowerInfer-2~\cite{xue2024powerinfer2},
Ripple~\cite{wang2024ripple}, and ActiveFlow~\cite{jia2025activeflow}, stream weights,
activations, or experts from off-package NVMe or universal flash storage (UFS) through the host storage stack, relying on the file system, page cache, or standard storage-class \gls{FTL}. 
Near- and in-storage inference systems, such as RecSSD~\cite{wilkening2021recssd}, DeepStore~\cite{mailthody2019deepstore}, Cognitive SSD~\cite{liang2019cognitive}, RM-SSD~\cite{sun2022rm}, ECSSD~\cite{li2023ecssd}, InstInfer~\cite{pan2025instattention}, and INF$^{2}$~\cite{jang2025inf}, push computation or semantic indexing into the storage device to amortize random-access overhead. 
Unified-memory and chiplet systems, such as G10~\cite{zhang2023g10} and
Cambricon-LLM~\cite{yu2024cambricon}, manage flash as part of a larger accelerator-visible address space or integrate NAND flash chiplets near the accelerator. }
\prop differs from these systems in three ways:
\li~the non-volatile tier is on-package \gls{HBF} reached through the \gls{HBM}-side \gls{D2D} link rather than off-package storage; compute remains on the GPU rather than moving into the storage device; and 
\lii~the storage-class \gls{FTL} is replaced with a burst-granular read-only translation table on the \gls{HBF} base die, eliminating OS-mediated page faults, page-granular metadata, and write-path machinery from the foreground inference path.}
\section{Conclusion}
\label{sec:conclusion}

\gf{\gfhpca{We introduce \prop, a workload-driven \gls{HBF} substrate
that specializes the \gls{HBF} access path for the burst-granular,
read-only behavior of \gls{LLM} inference weights through three
mechanisms: a hardware burst-buffer controller that stages bursts in
the buffers already inside each \gls{HBF} die, phantom-plane refresh
that keeps NAND maintenance off the foreground read path, and a
read-only \gls{FTL} that reduces address translation to a compact
burst-granular table. We conclude that workload-specialized \gls{HBF}
control turns on-package flash into a practical capacity tier for frontier \gls{LLM} inference. 
}}

\balance 
\bibliographystyle{IEEEtran}
\bibliography{refs}

\end{document}